\documentclass[reprint,aps,prx,twocolumn,amsmath,amssymb,floatfix,footinbib,bibnotes]{revtex4-1}

\pdfoutput=1

\usepackage[colorlinks=true,citecolor=blue,linkcolor=magenta]{hyperref}
\usepackage{amsmath}
\usepackage{graphicx}
\usepackage{dsfont}
\usepackage{changepage}
\usepackage{fancyhdr}
\usepackage{amsthm, amssymb}
\usepackage{float}
\usepackage{array}
\usepackage[usenames,dvipsnames]{color}

\newcommand{\be}{\begin{align}}
\newcommand{\ee}{\end{align}}

\def \be{\begin{equation}}
\def \ee{\end{equation}}
\def \ba{\begin{array}}
\def \ea{\end{array}}
\def \bea{\begin{eqnarray}}
\def \eea{\end{eqnarray}}
\def \nn{\nonumber}
\def \W{{\Omega}}
\def \e{{\epsilon}}
\def \ve{{\varepsilon}}
\def \l{{\lambda}}
\def \L{{\Lambda}}

\def \t{{\theta}}
\def \b{{\beta}}
\def \g{{\gamma}}

\def \w{{\omega}}
\def \s{{\sigma}}

\def \e{{\epsilon}}

\def \G{{\Gamma}}

\def \ba{\begin{align*}}
\def \ea{\end{align*}}

\newcounter{indice}

\def \mrm{\mathrm}

\def \bs{\boldsymbol}

\begin{document}

\title{Synthesizing Coulombic superconductivity in van der Waals bilayers }

%\author{Valla Fatemi and Jonathan Ruhman \\
%{  Department of Physics, Massachusetts Institute of Technology, Cambridge, MA 02139 USA  }}
\author{Valla Fatemi}
\affiliation{Department of Physics, Massachusetts Institute of Technology, Cambridge, MA 02139 USA }
\author{Jonathan Ruhman}
\email{ruhman@mit.edu}
\affiliation{Department of Physics, Massachusetts Institute of Technology, Cambridge, MA 02139 USA }

\begin{abstract}
{Synthesizing a polarizable environment surrounding a low-dimensional metal to generate superconductivity is a simple theoretical idea that still awaits a convincing experimental realization.
The challenging requirements are satisfied in a metallic bilayer when the ratio between the Fermi velocities is small and both metals have a similar, low carrier density.
In this case, the slower electron gas acts as a retarded polarizable medium (a ``dielectric" environment) for the faster metal.
Here we show that this concept is naturally optimized for the case of an atomically thin bilayer consisting of a Dirac semimetal (e.g. graphene) placed in atomic-scale proximity to a doped semiconducting transition metal dichalcogenide (e.g. WSe$_2$).
The superconducting transition temperature that arises from the dynamically screened Coulomb repulsion is computed using the linearized Eliashberg equation. In the case of graphene on WSe$_2$, we find that $T_c$ can exceed 100 mK, and it increases further when the Dirac valley degeneracy is reduced.
Thus, we argue that suspended van der Waals bilayers are in a unique position to realize experimentally this long anticipated theoretical concept.
} \end{abstract}

\maketitle

\section{Introduction}

In 1964 Little~\cite{Little1964} argued that superconductivity can be synthesized in low-dimensional conductors by placing them in proximity to a highly polarizable medium which converts the Coulomb repulsion into an effective attraction.
The idea attracted much attention due to the prediction of exceptionally high transition temperatures and was developed in many directions~\cite{Ginzburg1964,Allender1973,Bardeen1978,MERLIN1990743,gutfreund1996prospects,malozovsky1996metal,Cotle2016,Kavokin2016,Hamo2016}.
Nonetheless, there are no convincing experimental realizations of this elegant theoretical idea; superconductivity has not been synthesized using a polar medium thus far.
%One possible exception could be considered the enhancement of superconductivity in single layers of FeSe on SrTiO3. However, we find this highly inconsistent with the fact that similar enhancements were observed on amorphous TiO2 and by chemical doping.

Naively, attractive Coulomb interactions are a promising route to high-temperature superconductivity. However, converting the full strength of Coulomb repulsion into attraction poses major challenges, even theoretically. The first is that high density metals screen the Coulomb interaction quite effectively, thus suppressing the coupling strength. Moreover, to be effective, the separation between the metal and the polarizable medium must be smaller than the interparticle distance.
Thus, the most promising approach for the conductor is to use semimetals or doped semiconductors in which particle densities are low enough such that the interparticle distance can be greater than a few interionic distances.
%For example, the bare Coulomb pseudo-potential in the s-wave channel considered by Anderson and Morel[?] is maximally 1/2, not such a strong coupling at all.
The second challenge is that a stable dielectric medium has a positive static permittivity. Therefore, instantaneous attraction can only be obtained due to the quantum effects of ``over-screening", as demonstrated by Ref.~\cite{Hamo2016}. In the absence of such effects, the attraction must be generated dynamically, \'a la Anderson and Morel~\cite{Anderson1962}, which also reduces its bare strength.

These two challenges are anti-cooperative.
On the one hand, low-density metals entail a small Fermi energy, reducing the upper limit to the superconducting gap.
On the other hand, dynamical generation of attraction is efficient only when the Fermi energy is much greater than the characteristic frequencies in the dielectric medium, which are comparable in practice.
For example, the Fermi energy scale in semimetals or semiconductors is typically 10s to 100s of meV -- the same scale as longitudinal optical phonons in most dielectric materials.

\begin{figure}[h!]
 \begin{center}
    \includegraphics[width=1\linewidth]{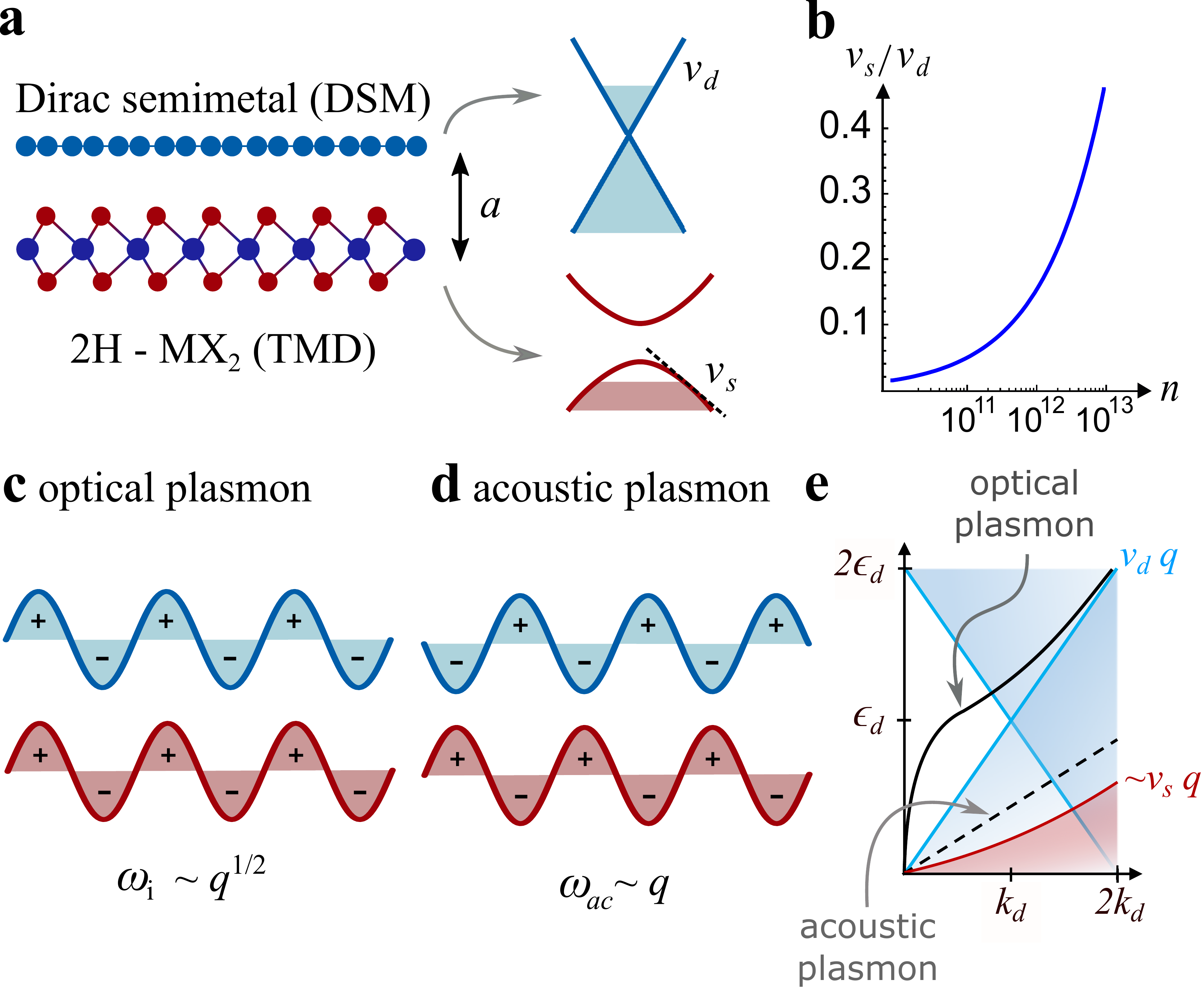}
 \end{center}
\caption{ The proposed setup for synthethesizing Coulombic superconductivity. $\bf{a}.$ A high velocity Dirac semimetal (DSM) is placed in proximity to a semiconducting transition metal dichalcogenide (TMD). The density of the two layers can be controlled independently using distant gate electrodes ~\cite{weitz2010broken}. $\bf{b}.$ As the density on both layers is tuned to zero, the ratio of their Fermi velocities also trends to zero (i.e. $ v_s/ v_d \ll 1$), where $v_d$ is the Dirac velocity and $v_s$ is the Fermi velocity of the TMD.  $\bf{c}.$ The optical plasmon is a collective mode of two propagating charge density waves, one on each layer, which are ``in-phase" with each other. $\bf{d}.$ The acoustic plasmon is a collective mode for which the two charge density waves are ``out-of-phase" with each other. As such, this mode is charge neutral overall, resulting in an acoustic mode. $\bf{e}.$ The schematic dispersion of the optical plasmon (solid) and the acoustic plasmon (dashed) superimposed on the particle-hole continuum of the Dirac semimetal (blue) and the TMD (red). The velocity of the acoustic mode scales as the geometric mean of the two Fermi velocities $\sim \sqrt{v_s v_d}$; thus it becomes very slow for low densities and can mediate superconductivity efficiently.   }
 \label{fig:setup}
\end{figure}

In this Letter, we propose a new route based on van der Waals (vdW) heterostructures~\cite{Geim2013van} in which a ``sluggish" conductor with slow carriers serves as the polarizable medium to mediate attraction.
Specifically, we propose a two-dimensional bilayer system consisting of a fast Dirac semimetal (DSM) layer, such as graphene, and a lightly doped semiconductor, such as Mo or W-based transition metal dichalcogenides (TMDs), to serve as the sluggish conductor, as schematically presented in Fig \ref{fig:setup}(a).
Using a numerical solution of the Eliashberg equation, we perform a careful study of the transition temperature for realistic devices.
From this analysis we conclude that Coulombic superconductivity is optimized by several parameters, for all of which our proposed vdW bilayers hold crucial advantages compared to traditional semiconductor double wells~\cite{Vakili2004,thakur1998electron}:

%
%We find that in a the required velocity ratio to obtain experimentally relevant transition temperatures can be achieved for the scenario of a DSM placed at atomic-scale separation to a parabolic band semiconductor. In brief, the critical temperature is optimized by the following parameters:

\noindent
(i) The ratio of the Fermi velocities of the two electronic systems must be very small, which is made possible in the vdW bilayer by the profound difference in electronic dispersion relations in the two layers [see Fig \ref{fig:setup} (b)].
In this limit the resultant retarded attractive interaction satisfies the conditions for BCS theory.

\noindent
(ii) The dielectric constant must be minimized in order to maximize the Coulomb interaction -- a surrounding dielectric strongly suppresses $T_c$.
vdW layers are unique in their ability to be suspended~\cite{bolotin2008ultrahigh,du2008approaching,weitz2010broken} and thus optimize the coupling strength, leading to experimentally relevant $T_c$ values (see Appendix~\ref{app:eps} for details).

\noindent
(iii) The distance between the electronic layers must be as small as possible in order to maximize their coupling. Also here, vdW heterostructures have a unique advantage by allowing for atomic-scale layer separations. We explore this in detail in Appendix \ref{app:layer_sep}.

\noindent
(iv) The strength of attraction is inversely proportional to the number of valleys in the fast layer. We analyze the transition temperature both for double-valley DSMs, such as graphene, as well as single valley DSMs. The latter obtains higher $T_c$.

\section{Pairing from interlayer acoustic plasmons }
To gain intuition for how the slow metallic layer serves to induce a retarded attractive interaction, we first discuss a limiting case in which we obtain analytic results. More rigorous numerical calculations estimating $T_c$ are presented in Section~\ref{sec:est_Tc}. For all calculations, we consider the situation of a two-dimensional DSM placed on top of a lightly doped semiconducting TMD. Hereafter, we label the two layers by $j = d,s$, corresponding to ``DSM" and ``semiconductor", respectively.

\subsection{Main concept}
To understand better how such a device can convert Coulomb repulsion into attraction let us consider the dynamical picture of electronic screening. In the bilayer configuration, the long ranged Coulomb interaction gives rise to {\it two} collective plasma modes - resonances of the screened electron-electron interactions. As in BCS theory each such mode can in principle mediate an effective attractive interaction~\cite{Takada1980} if it is retarded with respect to the Fermi energy.

The first mode is the standard {\it optical} plasmon, which is the ``in-phase" collective excitation of charge on the two layers [see Fig. \ref{fig:setup} (c)]. Unless a strong dielectric medium surrounds the device, this mode is higher than the Fermi energy except for very small momentum transfer. In addition to this mode there is an {\it acoustic} plasma mode, which describes collective ``out-of-phase" charge excitations of the two layers [see Fig. \ref{fig:setup} (d)]. Because the charge modulation on one layer is canceled by a negative charge on the other, this mode is neutral and thus acoustic [the dispersion of the two modes and the particle-hole continuum are schematically presented in Fig. \ref{fig:setup} (e)]. Thus, by adding the TMD layer, we effectively engineer an additional acoustic mode to the DSM, additive to the phonon modes.
The coupling of this mode to the electrons in the DSM is, however, of different origin and may therefore be larger.

The long-wavelength velocity of the acoustic plasma mode is set by the geometric mean of the Fermi velocities of the two layers, and it is only weakly damped by its coexistence with the particle-hole continuum of the DSM.
Specifically, for the limiting case of no separation between the layers ($a= 0$) and equal densities, this mode disperses as  $\w_{ac} =u_{ac}q$, where $u_{ac} =  u' - i\,u'' $, and
\be
u' =\sqrt{ v_s v_d \over 2\sqrt{G}}\;\;;\;\; u'' = \frac{v_s}{4\sqrt{G} } \label{disp}
\ee
where $G = g_\s^d g_v^d / g_\s^s g_v^s$ is the ratio of band degeneracies, where $g_\s^j$ and $g_v^j$ are the spin and valley degeneracies in layer $j$, respectively (see Appendix \ref{app:model} for details). For a DSM dispersing at $v_d = 10^6$ m/s (equivalent to graphene) and a TMD with effective mass $m_s = 0.5\, m_e$ one can easily reach the limit where there are orders of magnitude between the Fermi velocity of the DSM $v_s$ and that of the TMD by tuning to the low density limit [see Fig.~\ref{fig:setup}.(b)].
Thus, the BCS limit, in which the characteristic frequency of the acoustic mode is orders of magnitude smaller than the Fermi energy, is in reach. Moreover, unlike previous proposals based on parabolic bands~\cite{Pashitskii1969,frohlich1968superconductivity,RADHAKRISHNAN1965247,Takada1980,entin1984acoustic,Garland,Canright1989,thakur1998electron}, here the velocity ratio can be tuned over orders of magnitude by tuning the density.
We also note that this mode was considered in early discussions of superconductivity in highly doped graphene ~\cite{uchoa2007superconducting}. In Appendix ~\ref{app:AP_indic} we discuss possible methods to observe the acoustic plasma mode in the normal state.

As explained, a key aspect in our proposal is the great difference in Fermi velocities. This implies that the mode disperses inside the particle hole continuum of the DSM and is therefore damped, leading to a finite $u''$ in Eq.~\eqref{disp}.
It should be mentioned that the acoustic plasma mode is also often discussed in the context of double layers with similar velocities~\cite{Gabriele1988,Hwang2009,Rosario2012}. In that case it disperses outside of the particle-hole continuum and is undamped, which makes it much more visible much less effective for superconductivity.

Finally, we note that the acoustic mode velocity in Eq. ~\eqref{disp} does not depend on the Coulomb interaction. This is only an artifact of the $a = 0$ limit, where the only restoring force is the quantum compressibility of the gases. For any finite layer separation $a\ne0$ the velocity will also depend on the parameters of the Coulomb interaction (see Appendix~\ref{app:layer_sep}).

\subsection{The acoustic plasmon approximation}
For concreteness, let us consider the limit of equal density in the two layers $k_d = k_s/\sqrt{G}$, where $k_d$ ($k_s$) is the Fermi momentum in the DSM (semiconductor) layer.
The acoustic plasma mode described above separates two distinct regimes describing the Coulomb interaction within the DSM: at frequencies $\w > \w_{ac}$ the TMD is too slow to respond and does not participate in screening of the Coulomb interaction, whereas for $\w<\w_{ac}$, the TMD adds to the total screening and suppresses the interaction by a significant amount. By taking an approximation in the vicinity of the acoustic plasma mode (see Appendix ~\ref{app:APA} for details), we see that this manifests directly within the form of the Coulomb interaction at these two limits of high and low frequency:
\begin{align}
 &V_\infty(q)  =  {2\pi e^2/\ve  \over q + Q_d}\,,  \label{Vinf}\\
 &V_0(q) =  {2\pi e^2 /\ve  \over q + Q_d+Q_s} \, , \label{V0}
\end{align}
where $Q_j = 2g_\s^j g_v^j\pi e^2 \nu_j/\ve$ is the Thomas-Fermi wavevector of layer $j$, $\ve$ is the dielectric constant and $\nu_j = k_j/2\pi v_j$ is the density of states per species.
Indeed, one can see that the low-frequency interaction Eq. ~\eqref{V0} has added screening by the TMD accounted for by its Thomas-Fermi wave-vector, as compared with the high-frequency case Eq. ~\eqref{Vinf} where it is absent. The difference between these two limits, $V_\infty(q) - V_0(q) $, is the attraction strength generated by the TMD layer at a given $q$ (for an equivalent scenario using polar optical phonons see Ref.~\cite{Gurevich1962}). Note that both terms are positive (e.g. the Coulomb interaction is still repulsive at all frequencies), so this attraction strength is a \emph{relative} measure. To obtain effective attraction at low energy the high frequency repulsion must be screened in the standard manner~\cite{Anderson1962} (see next subsection).

The above limiting forms for the interaction can be connected by inspecting the Coulomb interaction in the vicinity of the acoustic mode, where it takes the form
\be
V_{ac} (\w, q) = {2 \pi e^2/\ve  \over  q + Q_d}{\left[1-\g (q){ (u q)^2 \over  (u q)^2 -\w ^2 } \right]  }\,. \label{acoustic}
\ee
Here $\g (q) = {Q_s / (q + Q_d + Q_s)}$ interpolates between the asymptotic behavior at high and low frequency [Equations \eqref{Vinf}  and \eqref{V0}]. Eq. \eqref{acoustic}, as an approximation of the full interaction, neglects the dynamics of the polarization of the DSM (including the optical plasmon); henceforth, this approximation will be referred to as the {\it acoustic plasmon approximation}.  Eq.~\eqref{acoustic} has been studied extensively in the context of multiband metals in which two bands with very different velocities are simultaneously occupied (see for example Refs.~\cite{pines1956electron,bennacer1989acoustic,bennacer1989calculated,chudzinski2011collective}). It also has the same form as the well known phonon mediated interaction in the classic theory of superconductivity~\cite{de2018superconductivity}. As a result, the acoustic plasmon has been proposed as a candidate mechanism for superconductivity by many authors in different contexts~\cite{Pashitskii1969,frohlich1968superconductivity,RADHAKRISHNAN1965247,Takada1980,entin1984acoustic,Garland,Canright1989,thakur1998electron,Ruhman2017}.

Inspecting Eqs.~(\ref{Vinf}-\ref{acoustic}), we find that the attraction strength, given by $V_\infty(q) - V_0(q) $, becomes stronger as the ratio between the Thomas-Fermi momenta grows and the velocity ratio decreases. In this limit the velocity of the mode~\eqref{disp} also becomes highly retarded.
The unique feature of the DSM-semiconductor bilayer, which makes it advantageous over previous proposals, is that the velocity ratio can be tuned and becomes infinite in the zero-density limit. Thus, the conditions for superconductivity are naturally optimized at low density.

\subsection{Analytic calculation of $T_c$ within the acoustic plasmon approximation}

Within BCS theory the $T_c$ resulting from \eqref{acoustic} is determined by three parameters -- the strength of the Coulomb pseudo-potential $\mu$, the attraction strength $\lambda$, and the bandwidth of the acoustic mode $\Theta_{ac}$ (in phonon superconductivity this is the Debye frequency) -- in the familiar form ${k_B T_c } \approx \Theta_{ac} \exp\left[{- {1\over \l - \mu^*}}\right]$.
%
%The variation between these two limits as a function of frequency occurs on the scale of the acoustic plasmon $\w_{ac} = u' q$, which is retarded with respect to $\e_d$ and defines the effective ``Debye frequency" for this mode when $q \sim 2k_F$.
To estimate $\mu$ and $\l$, let us assume pairing in the s-wave channel. When the Bloch bands are trivial these parameters are given by averaging the interaction over the Fermi surface~\cite{Margine2013}. In a DSM there is an additional coherence factor ${1\over 2}\left(1+\cos \t_{\bs k,\bs k'} \right)$~\cite{Ruhman2017}, where $\bs k $ and $\bs k'$ are the incoming and outgoing momenta. Taking this factor into account, the bare Coulomb repulsion is given by
\be
\mu = {\nu_d \over 4\pi  } \int_{-\pi}^{\pi }d\t \left(1 + \cos \t \right) V_{\infty}(q) \label{mu}
\ee
and the zero frequency attraction is
\be
\l =  {\nu_d \over 4\pi  } \int_{-\pi}^{\pi }d\t \left(1 + \cos \t \right) \left[V_{\infty}(q)-V_0(q)\right] \label{lam}
\ee
where $q = k_F\sqrt{2-2\cos\t}$.
Finally, the bandwidth of the acoustic mode is estimated by $\Theta_{ac} \approx  u' k_F = \sqrt{\e_d \e_s/G}$.

The ratio $x \equiv \w_{0}/\e_d = \sqrt{\e_s \over \e_dG}$ quantifies the retardation and therefore controls the logarithmic screening of the instantaneous Coulomb repulsion \eqref{mu} ~\cite{de2018}
\be
\mu^* = {\mu\over 1- \mu \log x}\,.\  \label{mustar}
\ee
 Performing the integrals \eqref{mu} and \eqref{lam} we find that
 \be
 \l = {\mu \over 1 + {2 G x^2  } }   \,.\label{lamstar}
 \ee
 After rewriting the bandwidth as $\Theta_{ac} = 2 m_s v_d^2 x^3$, we come to our estimate for the transition temperature within the acoustic plasmon approximation:
 \be
k_B T_c \approx {2 m_s v_d^2 x^3 } \exp\left[ {-{1 \over {\mu \over 1+ 2G x^2} - {\mu\over 1- {\mu} \log x } }}\right]. \label{Tc}
\ee
The strength of the bare Coulomb repulsion \eqref{mu} also sets the scale for the attraction \eqref{lamstar} because they are both of the same origin.

\begin{figure}[t]
 \begin{center}
    \includegraphics[width=1\linewidth]{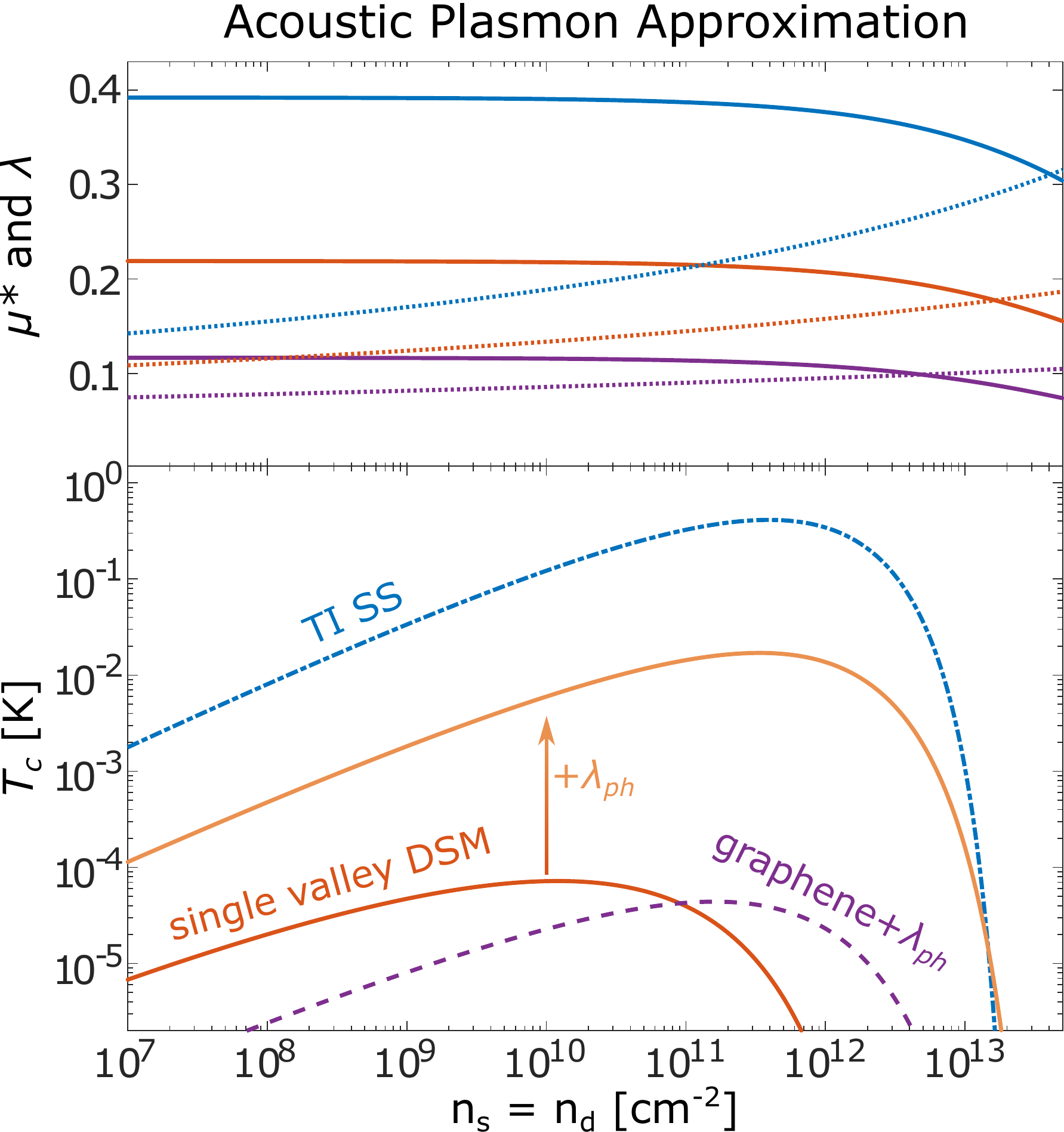}
 \end{center}
\caption{Results of the estimation of $T_c$ using only the attraction from the acoustic plasmon.
Top: the screened Coulomb repulsion $\mu^*$ \eqref{mustar} (dot-dashed lines) and the acoustic plasmon attraction $\l$ \eqref{lamstar} (solid lines) vs. density $n_s = n_d$ for different DSM degeneracies, indicated on by color on the lines in the lower panel.
Bottom: The corresponding transition temperature \eqref{Tc} vs. density for the different DSM cases.
An additional attraction coming from the phonons $\l_{ph} = 0.05$ \cite{Sun2014} is included in the graphene case, and we also show its impact on the single-valley DSM.  }
 \label{fig:Tc_ana}
\end{figure}

In Fig. \ref{fig:Tc_ana} we plot Eq.~\eqref{Tc} as a function of the density ($n=\frac{g_{d}}{\pi}m^{2}v_{d}^{2}x^{4}$) for different values of the DSM degeneracy: $g_\s^d g_v^d=1$ for the topological insulator (TI) surface state, $g_\s^d g_v^d=2$ for a spin-degenerate single-valley DSM, and $g_\s^d g_v^d=4$ for graphene.
This quantity sets the strength of the bare Coulomb repulsion $\mu \lesssim 1/2g_\s^d g_v^d$.
We also set realistic estimates for the mass in a TMD monolayer, $m = 0.5\,m_e$ ~\cite{Babak2016,Gustafsson2017arxiv}, its degeneracy $g_\s^s g_v^s = 2$ (corresponding to hole doping), and the velocity in existing DSMs $v_d =10^6 \,\mrm{m/s} $.

We expect the actual $T_c$ to be higher than estimated here due to three main factors:

\noindent
(i) So far we have neglected the dynamical part of the DSM polarization. Once it is taken into account there is also a positive contribution from the optical plasmon that increases $T_c$ into a measurable range (see Section III).

\noindent
(ii) Phonons in the DSM contribute to pairing in addition to the plasmonic modes. For example, in graphene the overall attraction due to phonons was estimated to be $\l_{ph} \approx 0.05$ using first principles calculations \cite{Sun2014}, which is a value typically insufficient to generate superconductivity. The phononic and plasmonic contributions are cooperative, and the total attraction can be as much as the sum of the two contributions $\l+\l_{ph}$ if $\Theta_D \sim \Theta_{ac}$. To illustrate that this cooperation can have a substantial impact, we also plot the $T_c$ for the single- and double-valley DSMs with an added phonon contribution in Fig. \ref{fig:Tc_ana}, which increases $T_c$ by several orders of magnitude. A recent manuscript has explored this interplay for single-layer materials~\cite{roesner2018plasmonic}.

\noindent
(iii) In Refs. ~\cite{Canright1989,Savary2017} it was shown that Fermi liquid corrections to the compressibility enhance \eqref{mu}, resulting in $\mu = (1+F_0^s)/2g_\s^d g_v^d$, where $F_0^s$ is the 0th symmetric Landau parameter and is generally of order 1. This increases the overall scale of $\l-\mu^*$ and therefore $T_c$.

\section{Estimation of $T_c$}\label{sec:est_Tc}

\subsection{Accounting for the full interaction}

In the previous section we estimated transition temperatures \eqref{Tc} using the acoustic plasmon approximation.
In this section, we estimate $T_c$ by more carefully accounting for the full dynamical and spatial form of both polarization functions. We use the linearized Eliashberg equation~\cite{takada1992plasmon}
\be
\Phi(i\w,k) = - {T_c \over  \nu_d L^2}\sum_{\w',k'} {\G(i \w-i \w',k,k')\Phi(i\w',k') \over \w'^2 + v_d^2 \left( k' - k_d\right)^2} \label{eliashberg}
\ee
to compute $T_c$ numerically, where the vertex function is given by the angular average of the {\it full} interaction (again we assume s-wave pairing):
\be
\G(i\w,k,k') = {\nu_d\over 4\pi}\int_{-\pi} ^\pi d\t \left(1 + \cos \t \right)V_{dd}(i\w,q) \label{vertex}
\ee
Here $q = |\bs k - \bs k'|$, $\t$ is the angle between $\bs k$ and $\bs k'$, and $V_{dd}$ is the full RPA interaction in the DSM layer (see Eq. \ref{Vdd}).
Note that in these equations we have switched to Matsubara frequencies, as the numerical solution is much simpler on the imaginary axis.

\begin{figure}
 \begin{center}
    \includegraphics[width=1\linewidth]{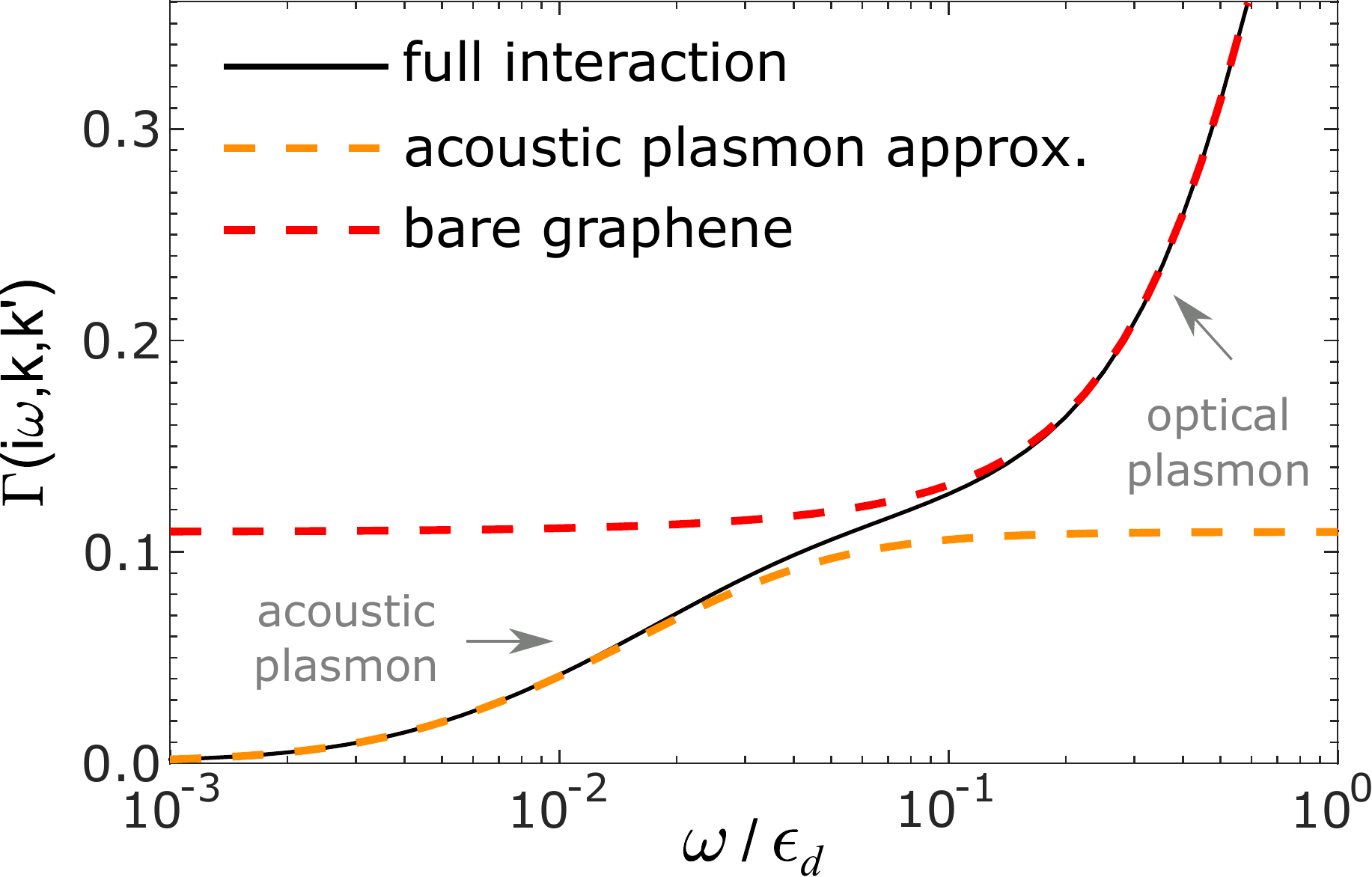}
 \end{center}
\caption{The vertex function for s-wave pairing, Eq.~\eqref{vertex}, as a function of Matsubara frequency for $k_d = k_s$, $k = 0.9 k_d$, $k'= 1.1 k_d$, $Q_d / k_d = 8.75$, and $m_s = 0.5m_e$.
The interaction decreases in two steps marked by grey arrows, corresponding to the dynamical contribution of the optical and acoustic plasma modes.
The red dashed curve denotes the vertex function without the semiconducting layer. The difference between these curves is the contribution of the acoustic plasmon, which leads to superconductivity. The dashed orange curve corresponds to the acoustic plasmon approximation, Eq.~\eqref{acoustic}, which does not account for the optical plasmon.}
 \label{fig:int}
\end{figure}

The vertex function \eqref{vertex} dictates $T_c$. Therefore, it will be instructive to understand its properties before we discuss the results for the calculation of $T_c$.
In Fig.~\ref{fig:int} we plot Eq.~\eqref{vertex} as a function of frequency (solid black) for $k = 0.9k_d$, $k' = 1.1k_d$, $g_v^d = 2$, $Q_d / k_d = 8.75$, $m_s = 0.5m_e$, $a = 0.5\mrm{nm}$, and $ n_d =n_s/2 = 10^{10}cm^{-2}$ ~\footnote{The value of $Q_d / k_d = 8.75$ corresponds to suspended graphene.}.
$\G(i\w,k,k')$ monotonically decreases in two steps, occurring at the acoustic and optical mode frequencies (in real frequency these steps are resonances). Each such step represents an attractive contribution to the interaction. These contributions add up in Eq.~\eqref{eliashberg}. However, since the overall interaction is repulsive at all frequencies it is crucial that the high frequency repulsion gets renormalized. The upper cutoff of the frequency summation is thus a crucial phenomenological parameter affecting the transition temperature.
We note that in our calculations the cutoff is never larger than the Fermi energy $\e_d$, and therefore the vertex function is plotted only up to that energy (see Appendix ~\ref{app:eliash} for details).

It is important to contrast the full interaction \eqref{vertex} with two limiting cases.
The first is the case of a bare DSM where there is no semiconducting layer, corresponding to the red dashed line in Fig.~\ref{fig:int}.
In this case the drop at lower frequency is absent, and the overall value of the repulsive interaction in the limit $\w\rightarrow 0$ is higher, which results in a weaker attraction at low energy.
The second limit is the acoustic plasmon approximation (\ref{acoustic}) obtained by taking the polarization of the DSM to be a constant $\Pi_d(\w,q) = g_v^d g_\s^d\nu_d$, represented by the dashed orange line in Fig.~\ref{fig:int}. In the range $\w > \w_{ac}$ the interaction goes to a constant.
Evidently, in this regime there are substantial deviations between this approximation (Eq. \eqref{acoustic}) and the full interaction.
These deviations contribute to elevate $T_c$ compared to Eq. \eqref{Tc} and are the effect of the optical plasmon.

\subsection{Results for $T_c$}

Let us now turn to the solution of $T_c$ obtained from solving Eq.~\eqref{eliashberg}. We account for a realistic separation between the center of the electronic wavefunctions in the vertical direction $a = 0.5 \mrm{nm}$~\cite{ma2011graphene}, and we use $Q_d = 4.3 k_d g_v^d$ as an estimate for the Thomas-Fermi wavevector of the DSM suspended in vacuum. The technical details of the numerical solution can be found in Refs.~\cite{takada1992plasmon,Ruhman2017} and in Appendix \ref{app:eliash}. In Fig. \ref{fig:Tc} the transition temperature $T_c$ as a function of the density in the semiconducting layer, $n_s$, is plotted for different values of the density in the DSM, $n_d$, and for the case of one and two valleys (i.e. $g_v^d =$ 1 and 2), corresponding to solid and dashed lines, respectively.

%We note that the numerical summation over frequency is truncated up to a cutoff $\W$, which is always taken to be smaller than the Fermi energy $\e_d$. The momentum is also integrated up to a cutoff around the Fermi energy $|k-k_d|< \L$, where $\L = \W/4v_d$. As always we discretize the momentum integration into $N_k$ points, distributed at a higher density close to $k = k_F$, and sum over a subset of the Matsubara frequencies~\cite{takada1992plasmon}. For more details see Appendix \ref{app:eliash}.

 %For these calculations we used $\W = 0.5\e_d $ for single valley and $\W = 0.8 \e_d $ for double valley DSMs. The latter cutoff is larger because the optical plasma frequency is greater by a factor of $\sqrt{2}$. (For a discussion on the dependence of $T_c$ on the cutoff see Appendix \ref{app:eliash}.)

\begin{figure}[t]
 \begin{center}
    \includegraphics[width=1\linewidth]{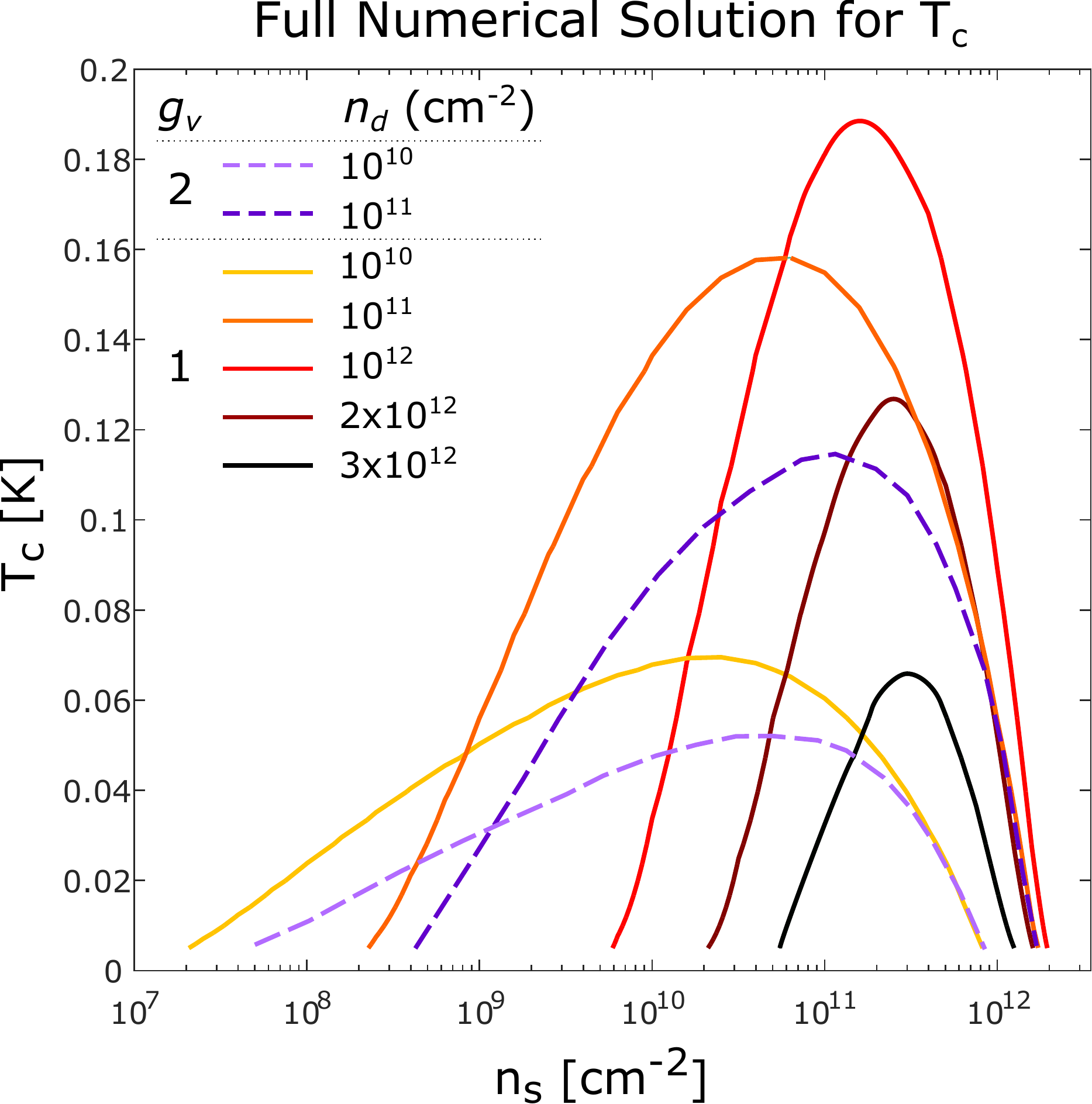}
 \end{center}
\caption{ The transition temperature obtained from the numerical solution of Eq.~\eqref{eliashberg} as a function of the density of the semiconducting layer $n_s$ for different values of the density of the Dirac semimetal $n_d$ (indicated in the legend).
Solid lines correspond to a single valley DSM, $g_v^d = 1$ and the dashed lines to double valley (i.e. graphene). All data for $g_\s^s g_\s^d = 2$ See Appendix~\ref{app:eliash} for technical details.  }
 \label{fig:Tc}
\end{figure}

 The transition temperature exhibits a dome shape, peaking at a non-universal value of the semiconductor density $n_s$. It is interesting to note that the domes are wide on a logarithmic scale, such that the two layers may have significantly different densities without strongly affecting $T_c$.
 The factors that dictate the suppression of $T_c$ for high and low $n_s$ are the following: in the limit $n_s \ll n_d$, $T_c$ diminishes because the semiconductor is unable to screen at wavevectors of order $k_d$. On the other hand, for $n_s \gg n_d$ the velocity ratio $v_s/v_d$ grows, thus harming the retardation condition. The peak is obtained by optimizing against these two parameters.

 Before proceeding we would like to make a few comments. Here we have assumed the dielectric screening by the environment is negligible, i.e. we took $\ve = 1$. $T_c$ is very sensitive to this parameter and estimates of its effect are discussed in Appendix \ref{app:eps}. Second, the layer separation distance $a$ also has a strong effect on $T_c$. Surprisingly, as we show in Appendix \ref{app:layer_sep}, it becomes influential even when the interparticle distance is two orders of magnitude greater than the layer separation.

 Finally, we comment that we have also used our calculation method to estimate the transition temperature in GaAs double quantum wells, where there is a large mass ratio between holes and electrons $m_h / m_e\approx 7$, which also quantifies the Fermi velocity ratio for equivalent layer density. This system was previously estimated to have a transition temperature of $100$ mK~\cite{thakur1998electron}, but superconductivity was never observed. Indeed, we found that $T_c$ is immeasurably low, in agreement with prior calculations using the RPA~\cite{Canright1989}. This is due to the large dielectric screening and large layer separation as well as the inability to tune the effective mass ratio, an important advantage of our proposed DSM-semiconductor bilayer. For more details see Appendix~\ref{app:GaAs}.

\section{Conclusions}
We have investigated the possibility that interlayer plasmons lead to superconductivity in vdW double layers. In particular we considered the scenario of a Dirac semimetal on top of a semiconducting transition metal dichalcogenide. Such a device has key advantages compared to previous proposals.

\noindent
(i) First, the different nature of the dispersions in these systems allows for arbitrary tuning of the velocity ratio. The velocity ratio plays an important role, both in the strength of the coupling and in the scale of retardation.

\noindent
(ii) Second, vdW devices can be suspended, placing them in an environment with the minimum possible dielectric constant and maximizing the Coulomb interaction strength.

\noindent
(iii) Finally, vdW heterostructures allow for atomic-scale interlayer separations. As we discuss in Appendix ~\ref{app:layer_sep}, $T_c$ drops with layer separation at a rate which is increased with coupling strength. Therefore, the ability to have atomic-scale interlayer separation is crucial in the limit of coupling strengths that lead to a measurable $T_c$.

All of these parameters affect the pairing strength dramatically. We calculated $T_c$ numerically using the linearized Eliashberg equation.
We found that in suspended devices of graphene on monolayer WSe$_2$ a maximal transition temperature exceeding 100 mK can be achieved within a realistic density range. Moreover, we showed that $T_c$ can be substantially enhanced by pre-existing electron-phonon interactions~\cite{Sun2014}, and we argued that accounting for corrections to the compressibility of the electronic liquid at short distances would enhance the overall coupling~\cite{Savary2017}. $T_c$ can be further enhanced if the number of valleys is reduced, but realistic material candidates for these cases have yet to be verified.

%$T_c$ can be further enhanced if the number of valleys is reduced. For example, in a single valley Dirac material (e.g. perhaps a single layer of ZrTe$_5$ ) the maximal temperature can almost reach $T_c = 200$ mK.
%However, it should be noted that such a semimetal can only occur accidentally, as there is no symmetry-protected single-valley Dirac semimetal in two-dimensions~\cite{Young2015}. Finally, the case of the topological insulator surface state would be ideal for maximizing $T_c$, but the typical dielectric constant of known materials is too high.

\section{Acknowledgments}
We are grateful to Avishai Benyamini, Justin Song, Patrick Lee, Marco Polini, and Hadar Steinberg for helpful discussions.
JR acknowledges a fellowship from the Gordon and Betty Moore Foundation under the EPiQS initiative (grant no. GBMF4303).

\appendix

\section{Model for the Bilayer System}\label{app:model}

\subsection{Model Basics}
We start from the dispersion Hamiltonian of the two layers, which is given by
\begin{align}
&H_d = \sum_k  \, \psi^\dag _{k} \left(v_d\,\bs k \cdot \bs \g-\e_d\right)  \psi_k \\ &H_s = \sum_k  \left({k^2 \over 2m_s}-\e_s\right) c^\dag _{k}  c_k\approx \sum_k  v_s\left(k-k_s\right) c^\dag _{k}  c_k\,,
\end{align}
where $\psi_k$ is a 4-component Dirac spinor in the basis of the Dirac matrices $\bs\g = (\g_1, \g_2)$ and $v_d$ and $\e_d$ are the Dirac velocity and Fermi energy of the DSM. Equivalently, $c_k$, $m_s$, $\e_s$, and $v_s$ are the field operators, mass, Fermi energy, and Fermi velocity of the semiconducting layer. $k_d$ and $k_s$ are the corresponding Fermi momenta. For all calculations we use the exact $H_s$.

\subsection{Electronic Polarizations and Coulomb Interactions }\label{app:pols}

The interactions between the layers generally have the form
\be
H_{I} =\sum_{\,ij = d,s} {n_i V_{ij} n_j   }
\ee
where $n_d$ and $n_s$ are the density operators of the two layers.
Within the random-phase-approximation (RPA) the matrix $V_{ij}$ assumes the form~\cite{Profumo2012}\
\begin{widetext}
\begin{align}
V_{ij}(i\w,q) ={1 \over A(i\w,q)}\label{full_int_mat}
\left(\begin{matrix}
V_q -\Pi_s(i\w,q)V_q^2\left(1 - e^{-2qa}\right) & V_q e^{-q a} \\
V_q e^{-q a} & V_q-\Pi_d(i\w,q)V_q^2 \left(1 - e^{-2qa}\right)
\end{matrix}\right)
\end{align}
where $a$ is the separation between the two layers [see Fig.~\ref{fig:setup} (a)] and the factor in the denominator is given by
\[A(i\w,q) = 1+ \Pi_d(i\w,q) \Pi_s(i\w,q)V_q^2\left(1 - e^{-2qa}\right)-\left[ \Pi_d(i\w,q)+\Pi_s(i\w,q) \right]V_q\,,\]
\end{widetext}
The bare Coulomb interaction is given by $V_q = 2\pi e^2 /\ve q$, where $\ve$ is a uniform dielectric constant. The polarization functions of the two layers are given by $\Pi_j(i\w,q) =g_\s^j g_v^j \nu_j P_j(i\w/\e_j, q/k_j) $, where $g_\s^j,\, g_v^j$ take into account any possible valley or spin degeneracies, $\nu_j =  k_j / 2\pi v_j $ is the density of states per species and the functions $P_j(i x,y)$ are well known~\cite{Barlas2007}
\begin{align}
& P_{s}(ix,y) =  -1+  \mrm{Im}\sqrt{{x^2 \over y^4} +{4 +  2i  x \over y^2 }-1}\nn\\
\label{pol} \\
& P_d(ix,y) =  -1-F\left({x}, y  \right)\left[\pi - 2 \,\mrm{Re}G\left( {2 + ix \over y}\right) \right] \nn
\end{align}
where $F(x,y) ={y^2 / 8 \sqrt{x^2  +y^2}}$ and $G(z) = \arcsin z+z\sqrt{1-z^2} $. Note that here, for the polarization of the DSM, we have taken both the interband and intraband contributions to the polarization.

Overall the interlayer interaction within the DSM, which we use in the Eliashberg equation \eqref{vertex}, is given by
\be
V_{dd}(i\w,q) = {V_q \over  { 1 - \Pi_s(i\w,q) V_q \over {1 -  \Pi_s(i\w,q) V_q\left[1-e^{-2 q a}\right]}  } -  \Pi_d(i\w,q)V_q } \label{Vdd}
\ee
In the limit of $a\rightarrow 0$ we obtain Eq.~\eqref{int}, and in the limit of $a \rightarrow \infty$ we restore the single layer result. Notice, however, that the exponential factor controlling the crossover between these two limits is $e^{2 q a}$. Thus, the appropriate length scale one must consider when comparing with the interparticle distance $l_d = 2\pi / k_d$ is not $a$ rather than $a' = 4\pi a$.

\subsection{Details of the Acoustic Plasmon Approximation }
\label{app:APA}
For simplicity, let us assume the density per flavor in each layer is equal, i.e. $k_d \approx k_s = k_F$. Furthermore, let us assume for a moment that the distance between the layers is much smaller than the interparticle distance, i.e. $k_F a \ll 1$. In this case we may assume that $U_q \approx V_q$ for all relevant momenta and thus the interaction \eqref{Vdd} assumes the simple form
\be
V_{dd}(\w,q) = {2\pi e^2 / \ve \over q-  \left[ Q_d P_d(\w,q) + Q_s P_s(\w,q) \right]}\label{int}
\ee
where $Q_j = 2\pi g_\s^j g_v^j e^2 \nu_j/\ve$ is the Thomas-Fermi wavelength of layer $j$.

The ratio between the Fermi velocities is given by $v_s / v_d = k_F / m_sv_d$, and thus approaches zero in the low density limit [see Fig.~\ref{fig:setup}(b) for realistic values]. Consequently, the frequency window $v_s q \ll \w \ll v_d q$ becomes parametrically large in the low density limit, which allows for the following approximations to the polarization functions:
\be
 P_d(\w,q) \approx  -1-{i\w \over v_d q}\;\;;\;\; P_s(\w,q) \approx   { v_s^2 q^2  \over 2\w ^2 } \label{pols_approx}
\ee
From these approximations to the polarization functions, we can find the pole that gives the acoustic plasma mode given by Eq. ~\eqref{disp} and reduce the full interaction to the acoustic plasmon approximation given in Eq. ~\eqref{acoustic}.

\section{Details of the numerical solution of the linearized Eliashberg equation}
\label{app:eliash}

\begin{figure}
 \begin{center}
    \includegraphics[width=1\linewidth]{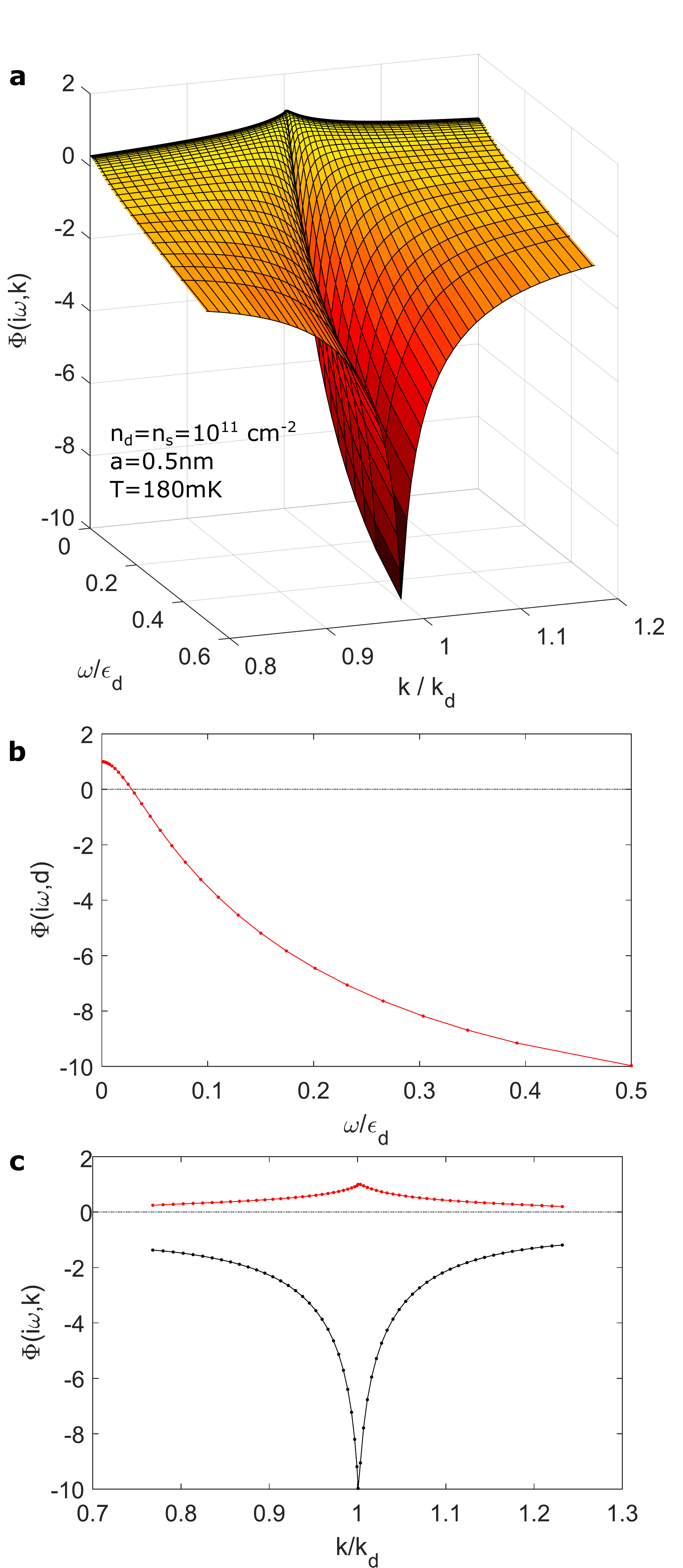}
 \end{center}
\caption{   (a) The eigenvector obtained from diagonalizing the kernel in Eq.~\eqref{eliashberg} for $T = 180$ mK, $n = n_d = n_s = 10^{11}\mrm{cm}^{-2}$, $g_v^d = 1$, $a = 0.5$, $N = 30$, $\b_k = 0.45$, $\b_\w = 4$, $\W = \e_d / 2$, $\L = \W/2v_d$, $v_d = 10^6$ m/s, and $m_s = 0.5 m_e$. (b) The frequency dependance of the eigenvector for $k = k_d$. (c) The momentum dependance for $\w = \pi T$ (red) and $\w = \W$ (black). }
 \label{fig:gap}
\end{figure}

The technique we have used here to solve for $T_c$ was first introduced by Takada~\cite{takada1992plasmon} and is detailed in Ref.~\cite{Ruhman2017}. Eq.~\eqref{eliashberg} is obtained from neglecting mass renormalization and dispersion corrections, and then linearizing the Eliashberg equation~\cite{Eliashberg1960,Margine2013} for the order parameter. Then the sum over frequencies and momenta is truncated at the cutoffs $\W$ and $\L = \W / 2 v_d$, respectively. Here we always limit the frequency cutoff by the Fermi energy, i.e. $\W \leq \e_d$.

In the next step only a subset of Matsubara frequencies and momenta points are chosen. The number of such points is labeled by $N_\w$ and $N_k$. We set $N_\w = N_k /2 = N$. Finally, the distribution of points is taken to be denser near the first Matsubara frequency $\Theta_{ac} = \pi T$ and the Fermi surface $k = k_d$. We choose an algebraically diverging density of points $p_\w(\w) = |\w|^{-\b_{\w}} $ and $p_k(k) = |k-k_d|^{-\b_{k}} $. For all simulations we take $\b_\w = 4$ and $\b_k = 0.45$.

$T_c$ is then obtained by seeking the value of $T$, to which the kernel on the r.h.s. of Eq.~\eqref{eliashberg} has a unity eigenvalue. Note that this must be the largest (positive) eigenvalue. The corresponding eigenvector at $T = 180$ mK, $n = n_d = n_s = 10^{11}\mrm{cm}^{-2}$, $g_v^d = 1$ and $a = 0.5$ nm is plotted in Fig.~\ref{fig:gap}(a). Note that this represents the gap at $T = T_c$ such that the overall scale of the gap is arbitrary [we chose to normalize by $\Phi (0,0)$].

The dependence on frequency is plotted in Fig.~\ref{fig:gap}(b) for $k = k_d$. As in the standard theory of superconductivity we find that the gap function changes sign at the frequency of the acoustic plasmon mode (in this case $\w_{ac}\sim  0.025\e_d$). In Fig.~\ref{fig:gap}(c) the momentum dependance of the gap function is plotted for two frequencies: $\w = \pi T$ (red) and $\w = \W$ (black). The sharp feature near $k = k_d$ is captured by the appropriate density of points controlled by $\b _k$.

\begin{figure}
 \begin{center}
    \includegraphics[width=1\linewidth]{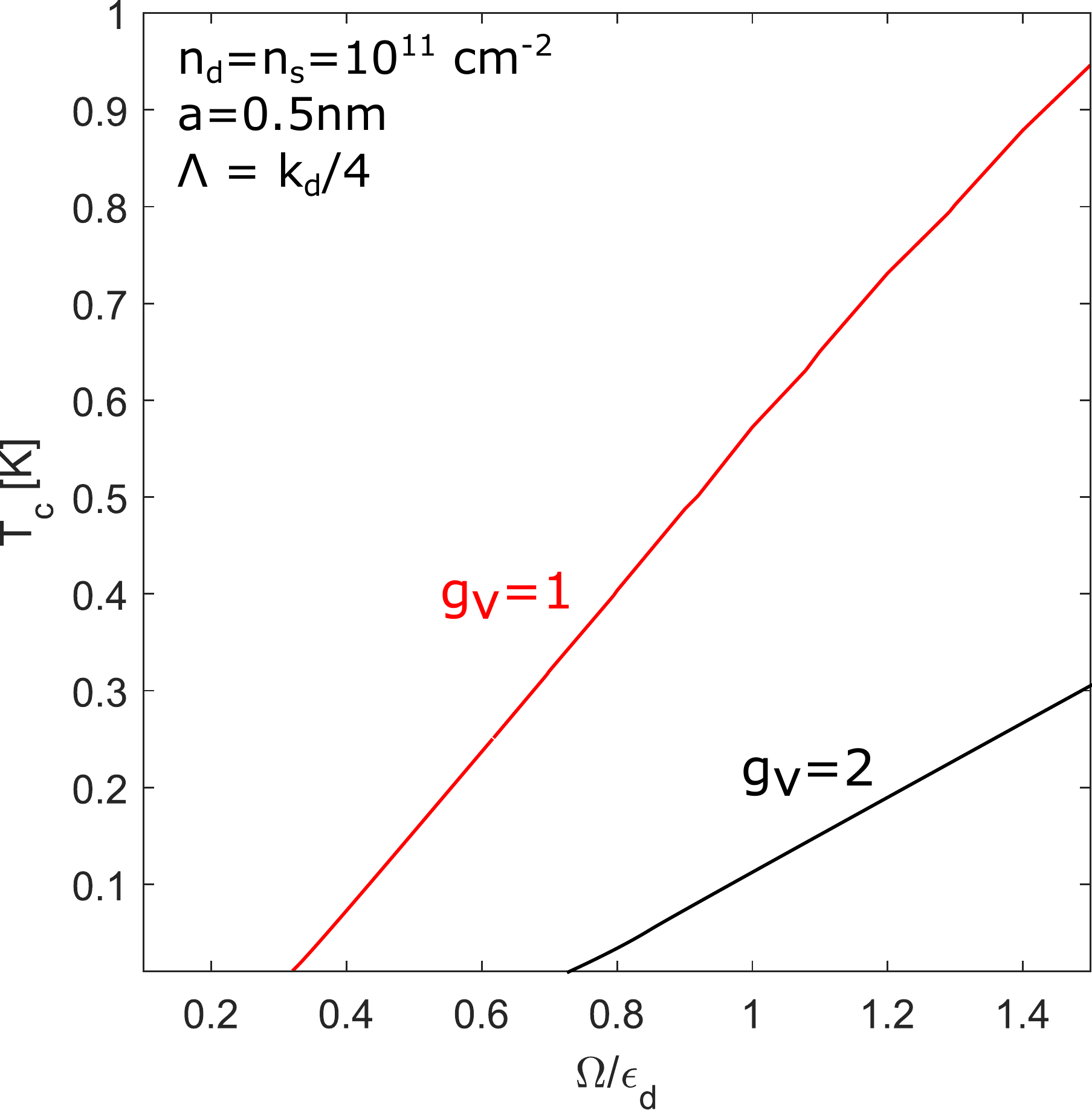}
 \end{center}
\caption{ The dependence of $T_c$ on the frequency cutoff $\W$. }
 \label{fig:vsOmega}
\end{figure}

The dependence of $T_c$ on the frequency cutoff is presented in Fig.~\ref{fig:vsOmega}. $T_c$ is found to be mostly linear in the cutoff $\W$ in the range of interest. We point out that this is not inconsistent with the standard formula for $T_c$ in the weak coupling limit (e.g. Eq.~\ref{Tc}) where the cutoff $\W$ tunes the parameter $\mu^*$~\cite{Anderson1962,Margine2013}. As such, the cutoff must be considered as a phenomenological parameter. Since Eliashberg theory, which is based on the sum over Gor'kov ladder diagrams, is justified only at low energy compared to $\e_d$ we insist that the cutoff must not be larger than $\e_d$. In doing so, we differ from previous studies (e.g. Refs. \cite{Takada1980,Canright1989}) by taking this more conservative approach.

\begin{figure}
 \begin{center}
    \includegraphics[width=1\linewidth]{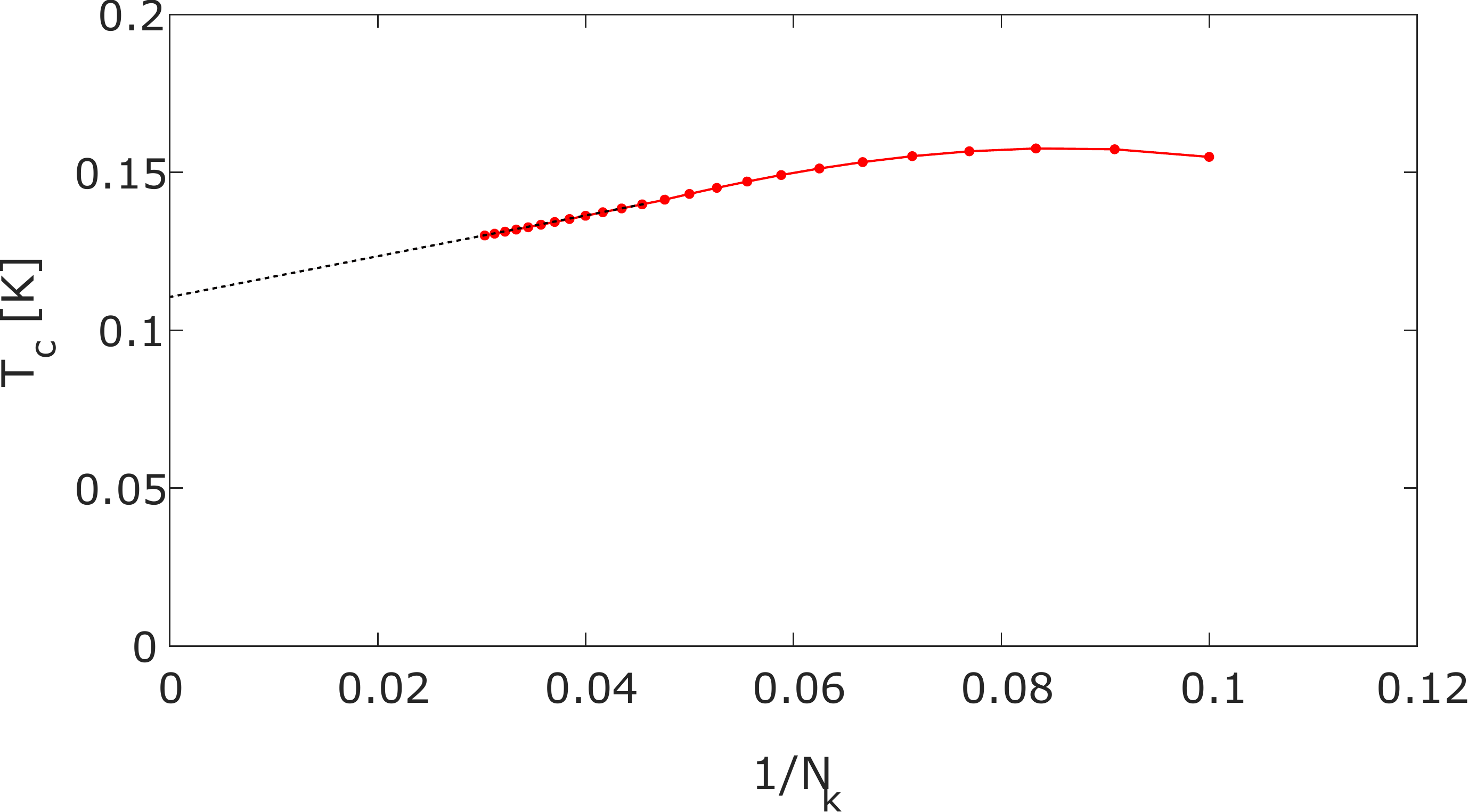}
 \end{center}
\caption{ The dependence of $T_c$ on the number of grid $1/N_k$ for $N_\w =20$, $\b_k = 0.55$, $\b_\w = 4$, $\W = 0.5 \e_d$, $\L = 0.2 k_d$. The dashed black line is the interpolation to the continuum limit $N_k \rightarrow \infty$.}
 \label{fig:vsN}
\end{figure}

The dependence of the transition temperature with the number of points in momentum space $N_k$ is plotted in Fig.~\ref{fig:vsN}.
As can be seen the transition temperature decreases slowly towards the thermodynamic limit but clearly converges to a finite value.

We also note that at sufficiently low temperatures the Eliashberg equation predicts a small but finite $T_c$ even in the case of a single two-dimensional metallic layer due to the optical plasmon, as pointed out by Takada long ago~\cite{Takada1978}. In the whole parameter range we have studied here this temperature is lower than 5 mK, which is the minimal temperature in our numerical simulation.
%As pointed out recently by Ref.\cite{Ruhman2017} this solution can be very sensitive to the mesh density near $k = k_F$, which can be set by $N_k$ and $\b_k$.

\section{Effect of dielectric screening}\label{app:eps}
\begin{figure}
 \begin{center}
    \includegraphics[width=1\linewidth]{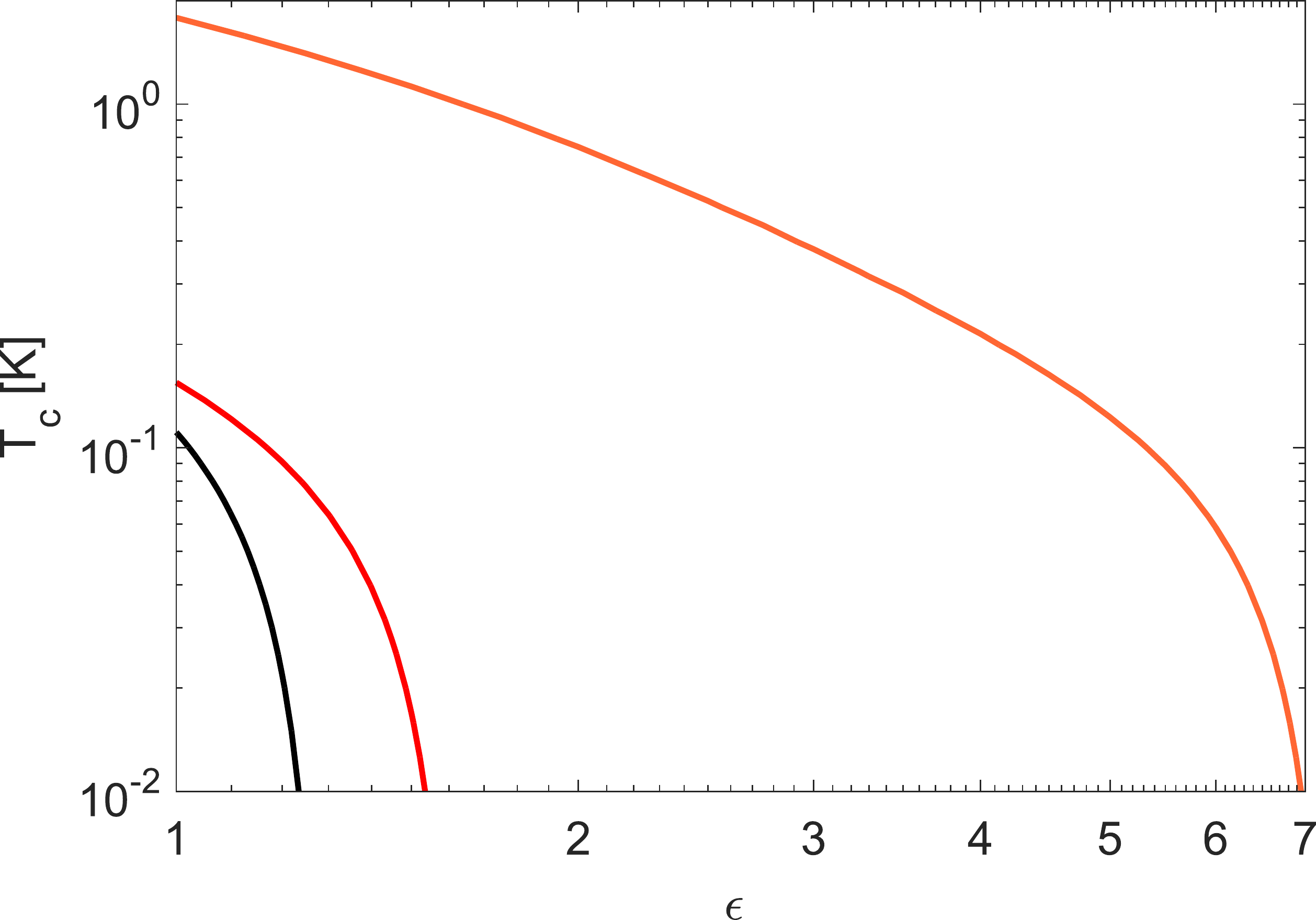}
 \end{center}
\caption{ The transition temperature vs the dielectric constant $\ve$ calculated numerically from Eq.~\eqref{eliashberg} for three different cases: double valley DSM - black, single valley DSM - red and for surface states of a topological insulator - orange. The frequency cutoff $\W$ for these three cases are $\e_d$, $0.5 \e_d$ and $0.32 \e_d$, respectively. All other parameters are the same as in Fig.~\ref{fig:Tc}. }
 \label{fig:Tc_vs_eps}
\end{figure}

In the main text we have assumed a vacuum dielectric constant of $\ve  =1$. In this section we compute the effects of finite dielectric screening.
In Fig.~\ref{fig:Tc_vs_eps} we plot $T_c$ vs. the dielectric constant $\ve$ for $g_v^d = 2,\, 1$, and the case of a 3D topological insulator (TI) surface state ($g_v^d = g_\s ^d = 1$). $T_c$ is much larger in the last case: $T_c = 1.8 $ K for $\ve = 1$ and becomes immeasurably small at around $\ve = 7$.
This plot shows that superconductivity is extremely sensitive to the dielectric environment surrounding the device.
It is however, important to note that in the case where only a half plane is polarizable this value corresponds to half of the bulk value of $\ve$. Finally, we note that the large $T_c$ in the TI surface states naively makes them the most promising candidates to realize plasmonic superconductivity. However, all known realizations of the 3D TIs also have a huge dielectric constant.

\section{Effect of layer separation}\label{app:layer_sep}
\begin{figure}
 \begin{center}
    \includegraphics[width=1\linewidth]{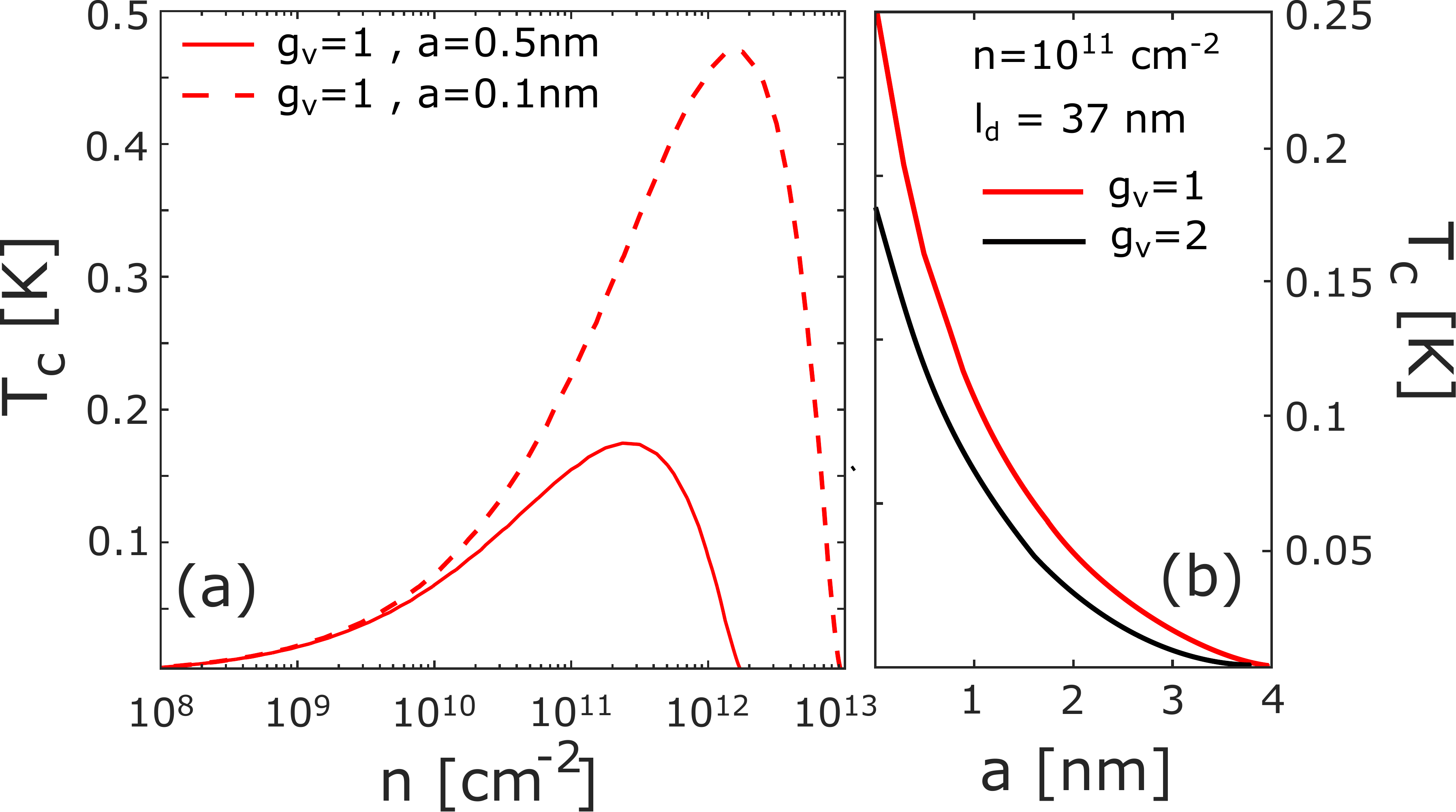}
 \end{center}
\caption{  (a) $T_c$ vs density for the case of $n = n_d = n_s$ for the case of a single valley DSM $g_v = 1$. Solid lines correspond to $a = 0.5$ nm and dashed lines to $a = 0.1$ nm. (b) $T_c$ as a function of the layer separation $a$. Red curves correspond to single valley DSMs and black to double valley DSMs.  }
 \label{fig:Tc_sweep_v2}
\end{figure}

In the main text we have assumed the minimum possible layer separation $a = 0.5$ nm. In this section we compute the effects of modifying the layer separation.
 In Fig.~\ref{fig:Tc_sweep_v2}(a) we plot $T_c$ for single valley DSMs as a function of density (solid curves) for the case of $n = n_d = n_s$, where all other parameters are as in Fig.~\eqref{fig:Tc}.
 As before we find that $T_c$ exhibits a dome as a function of total density.
 At extremely low density the two curves coincide signaling that the effects of modifying the layer separation from $0.1$ nm to $0.5$ nm vanishes when the interparticle distance is big enough. In this limit $T_c$ drops because the overall scale for the acoustic plasmon $\w_{ac} \sim \sqrt{\e_d \e_s}$ decreases.
 On the other hand, the main factor that suppresses $T_c$ in the high density limit is the finite separation between the layers, $a$.
 This can be seen by comparing the $a=0.5$ nm (solid) curve to the $a = 0.1 \mrm{nm}$ (dashed). The latter has a finite $T_c$ up to much higher density values.

 To further clarify this point we plot $T_c$ as a function of the layer separation, $a$, for $n_d = 10^{11}\mrm{cm}^{-2}$ in Fig.~\ref{fig:Tc_sweep_v2}(b). Interestingly, $T_c$ is strongly affected by layer separation on the scale despite the very large interparticle separation ($l_d \sim 37$ nm). Thus, to avoid significant suppression of the transition temperature the layer separation must be smaller than the interparticle distance $l_d$ by almost two orders of magnitude.

 We can understand the high sensitivity to layer separation by inspecting the coupling constant for the interaction \eqref{Vdd}, finding that it is reduced linearly in $a$: $V_{dd}/V_q \approx 1 - 2 Q_d a$. Note that, for suspended graphene, $2 Q_d a \sim 15.5 k_d a$. Thus, the more stringent requirement $k_d a \ll 0.1$ is apparent, indicating the interplay of these length-scales with the Coulomb interaction. This also implies that as the coupling becomes stronger the sensitivity to layer separation becomes greater.

 It was also mentioned in the main text that the velocity of the acoustic mode can acquire additional stiffness due to Coulomb interactions at finite layer separation $a\neq0$. This can be quantified by solving for the pole of the acoustic plasma mode for small $qa\sim k_F a<1$. In this limit, we find
 \be
 u_{ac} \approx \sqrt{ \frac{v_s v_d}{2} \left( 1 + 2 Q_d a \right) }
 \ee
 So again the layer separation plays an important role, although less so than in the coupling constant.

\section{The case of GaAs double wells}
\label{app:GaAs}
Superconductivity from interlayer plasmons has been proposed in the past in the context of electron-hole double layer quantum wells in GaAs~\cite{Vakili2004}. A transition temperature of 100 mK was estimated~\cite{thakur1998electron}, but superconductivity was never observed experimentally. To make a comparison with these previous predictions, we checked the prediction of Eq. ~\eqref{eliashberg} with the same parameters as in Ref. ~\cite{thakur1998electron}; however, we did not find a superconducting instability for $T > 0.5$ mK.

For this calculation, we took two single valley parabolic bands with equal and opposite density of $n = 10^{10} \mrm{cm}^{-2}$, the masses are taken to be $m_e = 0.03$ and $m_h = 0.22$, $\ve = 15$ and $a = 15$ nm. Otherwise, we used the same numerical parameters that were used to generate Fig.~\ref{fig:Tc}, except for the cutoff $\W $, which in this case we took to be higher, namely equal to the Fermi energy of the electron band (with the higher Fermi energy of the two).

$T_c$ is suppressed in the GaAs double wells mainly because the mass ratio, $m_h/ m_e \approx 7.1$ is not large enough, the dielectric environment strongly screens the interaction, and the quantum wells are relatively far apart. This highlights the several advantages of DSM-semiconductor double layers based on van der Waals materials: there is no theoretical bound on the effective mass ratio (or more accurately on the velocity ratio); the electronic system can be subjected to a more variable dielectric environment; and the interlayer separation can be taken to be much smaller.

\section{Non-superconducting indicators of the acoustic plasma mode} \label{app:AP_indic}
In the main text we investigated the possibility that the acoustic plasma mode in DSM-semiconductor bilayers leads to a superconducting instability. We would like to emphasize, however, that the experimental observation of this mode, even without superconductivity, is of fundamental interest. To the best of our knowledge such a mode was never observed in the limit of such a large velocity ratio, wherein the mode lies in the particle-hole continuum. Similar physics was found for surface plasmons on 3d metals \cite{diaconescu2007low,lundeberg2017tuning} as well as for photoexcited 3d semiconductors~\cite{Padmanabhan2014}. Moreover, it will be important to verify that the acoustic plasma mode exists in DSM-semiconductor bilayers in addition to the search for the superconductivity.

{\it Transport -- } The existence of a low energy acoustic mode was predicted to lead to phonon like scattering~\cite{chudzinski2011collective} and thus contribute to resistivity. In the case of two-dimensions this scattering mechanism should lead to a contribution $\rho_{ac} \propto (T/\Theta_{ac})^4$, where the energy scale $\Theta_{ac}$ is proportional to $\w_{ac}$ and is therefore density dependent. It is important to note that such a contribution can only appear if the momentum decay rate of the acoustic plasma mode to impurities is much greater than to electrons.
Another transport indicator of the acoustic plasmon is expected in the interlayer Coulomb drag signal. Here a distinctive non-monotonic temperature dependance has been predicted ~\cite{Flensberg1994}.

{\it Tunneling -- } We also expect that the acoustic plasmon can be measured using inelastic tunneling between the layers~\cite{de2017plasmon}.
Plasmonic spectroscopic signatures have been measured in GaAs quantum wells~\cite{jang2017full} and high quality tunneling data can be achieved by using van der Waals layers as a tunnel barrier~\cite{dvir2018spectroscopy}. It would also be interesting if the scanning near field probes could find a method to couple to this mode~\cite{fei2012gate,alonso2016acoustic}. However, it is important to note that because of the small layer separation $k_F a \ll 1$, the dipole moment of the mode is expected to be extremely small.

{\it Optics -- } The acoustic plasmon is essentially a longitudinal mode involving the relative charge oscillations on the two layers. As such it can be measured using Raman spectroscopy, as demonstrated in GaAs quantum wells~\cite{PhysRevB.53.11016,PhysRevB.57.R2065,PhysRevB.59.2095}.

{\it Plasmon - phonon interaction --} Finally, the acoustic plasmon is allowed to couple to other longitudinal waves, such as acoustic and optical phonons. At points where the dispersion branches cross, strong phonon-plasmon coupling is expected. However, it is important to note that even in the extremely dilute limit the velocity of the acoustic plasmon is expected to be larger than the acoustic phonon velocity in graphene.


\begin{thebibliography}{61}%
\makeatletter
\providecommand \@ifxundefined [1]{%
 \@ifx{#1\undefined}
}%
\providecommand \@ifnum [1]{%
 \ifnum #1\expandafter \@firstoftwo
 \else \expandafter \@secondoftwo
 \fi
}%
\providecommand \@ifx [1]{%
 \ifx #1\expandafter \@firstoftwo
 \else \expandafter \@secondoftwo
 \fi
}%
\providecommand \natexlab [1]{#1}%
\providecommand \enquote  [1]{``#1''}%
\providecommand \bibnamefont  [1]{#1}%
\providecommand \bibfnamefont [1]{#1}%
\providecommand \citenamefont [1]{#1}%
\providecommand \href@noop [0]{\@secondoftwo}%
\providecommand \href [0]{\begingroup \@sanitize@url \@href}%
\providecommand \@href[1]{\@@startlink{#1}\@@href}%
\providecommand \@@href[1]{\endgroup#1\@@endlink}%
\providecommand \@sanitize@url [0]{\catcode `\\12\catcode `\$12\catcode
  `\&12\catcode `\#12\catcode `\^12\catcode `\_12\catcode `\%12\relax}%
\providecommand \@@startlink[1]{}%
\providecommand \@@endlink[0]{}%
\providecommand \url  [0]{\begingroup\@sanitize@url \@url }%
\providecommand \@url [1]{\endgroup\@href {#1}{\urlprefix }}%
\providecommand \urlprefix  [0]{URL }%
\providecommand \Eprint [0]{\href }%
\providecommand \doibase [0]{http://dx.doi.org/}%
\providecommand \selectlanguage [0]{\@gobble}%
\providecommand \bibinfo  [0]{\@secondoftwo}%
\providecommand \bibfield  [0]{\@secondoftwo}%
\providecommand \translation [1]{[#1]}%
\providecommand \BibitemOpen [0]{}%
\providecommand \bibitemStop [0]{}%
\providecommand \bibitemNoStop [0]{.\EOS\space}%
\providecommand \EOS [0]{\spacefactor3000\relax}%
\providecommand \BibitemShut  [1]{\csname bibitem#1\endcsname}%
\let\auto@bib@innerbib\@empty
%</preamble>
\bibitem [{\citenamefont {Little}(1964)}]{Little1964}%
  \BibitemOpen
  \bibfield  {author} {\bibinfo {author} {\bibfnamefont {W.~A.}\ \bibnamefont
  {Little}},\ }\href {\doibase 10.1103/PhysRev.134.A1416} {\bibfield  {journal}
  {\bibinfo  {journal} {Physical Review}\ }\textbf {\bibinfo {volume} {134}}
  (\bibinfo {year} {1964}),\ 10.1103/PhysRev.134.A1416}\BibitemShut {NoStop}%
\bibitem [{\citenamefont {Ginzburg}(1964)}]{Ginzburg1964}%
  \BibitemOpen
  \bibfield  {author} {\bibinfo {author} {\bibfnamefont {V.~L.}\ \bibnamefont
  {Ginzburg}},\ }\href {\doibase 10.1209/0295-5075/85/27009} {\bibfield
  {journal} {\bibinfo  {journal} {Physics Letters}\ }\textbf {\bibinfo {volume}
  {13}},\ \bibinfo {pages} {101} (\bibinfo {year} {1964})},\ \Eprint
  {http://arxiv.org/abs/1412.0460} {arXiv:1412.0460} \BibitemShut {NoStop}%
\bibitem [{\citenamefont {Allender}\ \emph {et~al.}(1973)\citenamefont
  {Allender}, \citenamefont {Bray},\ and\ \citenamefont
  {Bardeen}}]{Allender1973}%
  \BibitemOpen
  \bibfield  {author} {\bibinfo {author} {\bibfnamefont {D.}~\bibnamefont
  {Allender}}, \bibinfo {author} {\bibfnamefont {J.}~\bibnamefont {Bray}}, \
  and\ \bibinfo {author} {\bibfnamefont {J.}~\bibnamefont {Bardeen}},\ }\href
  {\doibase 10.1103/PhysRevB.7.1020} {\bibfield  {journal} {\bibinfo  {journal}
  {Phys. Rev. B}\ }\textbf {\bibinfo {volume} {7}},\ \bibinfo {pages} {1020}
  (\bibinfo {year} {1973})}\BibitemShut {NoStop}%
\bibitem [{\citenamefont {Bardeen}(1978)}]{Bardeen1978}%
  \BibitemOpen
  \bibfield  {author} {\bibinfo {author} {\bibfnamefont {J.}~\bibnamefont
  {Bardeen}},\ }\href {\doibase 10.1016/0022-5088(78)90058-9} {\bibfield
  {journal} {\bibinfo  {journal} {Journal of The Less-Common Metals}\ }\textbf
  {\bibinfo {volume} {62}},\ \bibinfo {pages} {447} (\bibinfo {year}
  {1978})}\BibitemShut {NoStop}%
\bibitem [{\citenamefont {Merlin}(1990)}]{MERLIN1990743}%
  \BibitemOpen
  \bibfield  {author} {\bibinfo {author} {\bibfnamefont {R.}~\bibnamefont
  {Merlin}},\ }\href {\doibase https://doi.org/10.1016/0038-1098(90)90238-7}
  {\bibfield  {journal} {\bibinfo  {journal} {Solid State Communications}\
  }\textbf {\bibinfo {volume} {75}},\ \bibinfo {pages} {743 } (\bibinfo {year}
  {1990})}\BibitemShut {NoStop}%
\bibitem [{\citenamefont {Gutfreund}\ and\ \citenamefont
  {Little}(1996)}]{gutfreund1996prospects}%
  \BibitemOpen
  \bibfield  {author} {\bibinfo {author} {\bibfnamefont {H.}~\bibnamefont
  {Gutfreund}}\ and\ \bibinfo {author} {\bibfnamefont {W.}~\bibnamefont
  {Little}},\ }in\ \href@noop {} {\emph {\bibinfo {booktitle} {From
  High-Temperature Superconductivity to Microminiature Refrigeration}}}\
  (\bibinfo  {publisher} {Springer},\ \bibinfo {year} {1996})\ pp.\ \bibinfo
  {pages} {65--132}\BibitemShut {NoStop}%
\bibitem [{\citenamefont {Malozovsky}\ and\ \citenamefont
  {Fan}(1996)}]{malozovsky1996metal}%
  \BibitemOpen
  \bibfield  {author} {\bibinfo {author} {\bibfnamefont {Y.}~\bibnamefont
  {Malozovsky}}\ and\ \bibinfo {author} {\bibfnamefont {J.}~\bibnamefont
  {Fan}},\ }\href@noop {} {\bibfield  {journal} {\bibinfo  {journal}
  {Superconductor Science and Technology}\ }\textbf {\bibinfo {volume} {9}},\
  \bibinfo {pages} {622} (\bibinfo {year} {1996})}\BibitemShut {NoStop}%
\bibitem [{\citenamefont {Cotle\ifmmode~\mbox{\c{t}}\else \c{t}\fi{}}\ \emph
  {et~al.}(2016)\citenamefont {Cotle\ifmmode~\mbox{\c{t}}\else \c{t}\fi{}},
  \citenamefont {Zeytino\ifmmode~\check{g}\else \v{g}\fi{}lu}, \citenamefont
  {Sigrist}, \citenamefont {Demler},\ and\ \citenamefont
  {Imamo\ifmmode~\check{g}\else \v{g}\fi{}lu}}]{Cotle2016}%
  \BibitemOpen
  \bibfield  {author} {\bibinfo {author} {\bibfnamefont {O.}~\bibnamefont
  {Cotle\ifmmode~\mbox{\c{t}}\else \c{t}\fi{}}}, \bibinfo {author}
  {\bibfnamefont {S.}~\bibnamefont {Zeytino\ifmmode~\check{g}\else
  \v{g}\fi{}lu}}, \bibinfo {author} {\bibfnamefont {M.}~\bibnamefont
  {Sigrist}}, \bibinfo {author} {\bibfnamefont {E.}~\bibnamefont {Demler}}, \
  and\ \bibinfo {author} {\bibfnamefont {A.~m.~c.}\ \bibnamefont
  {Imamo\ifmmode~\check{g}\else \v{g}\fi{}lu}},\ }\href {\doibase
  10.1103/PhysRevB.93.054510} {\bibfield  {journal} {\bibinfo  {journal} {Phys.
  Rev. B}\ }\textbf {\bibinfo {volume} {93}},\ \bibinfo {pages} {054510}
  (\bibinfo {year} {2016})}\BibitemShut {NoStop}%
\bibitem [{\citenamefont {Kavokin}\ and\ \citenamefont
  {Lagoudakis}(2016)}]{Kavokin2016}%
  \BibitemOpen
  \bibfield  {author} {\bibinfo {author} {\bibfnamefont {A.}~\bibnamefont
  {Kavokin}}\ and\ \bibinfo {author} {\bibfnamefont {P.}~\bibnamefont
  {Lagoudakis}},\ }\href {\doibase 10.1038/nmat4646} {\bibfield  {journal}
  {\bibinfo  {journal} {Nature Publishing Group}\ }\textbf {\bibinfo {volume}
  {15}},\ \bibinfo {pages} {599} (\bibinfo {year} {2016})}\BibitemShut
  {NoStop}%
\bibitem [{\citenamefont {Hamo}\ \emph {et~al.}(2016)\citenamefont {Hamo},
  \citenamefont {Benyamini}, \citenamefont {Shapir}, \citenamefont {Khivrich},
  \citenamefont {Waissman}, \citenamefont {Kaasbjerg}, \citenamefont {Oreg},
  \citenamefont {{Von Oppen}},\ and\ \citenamefont {Ilani}}]{Hamo2016}%
  \BibitemOpen
  \bibfield  {author} {\bibinfo {author} {\bibfnamefont {A.}~\bibnamefont
  {Hamo}}, \bibinfo {author} {\bibfnamefont {A.}~\bibnamefont {Benyamini}},
  \bibinfo {author} {\bibfnamefont {I.}~\bibnamefont {Shapir}}, \bibinfo
  {author} {\bibfnamefont {I.}~\bibnamefont {Khivrich}}, \bibinfo {author}
  {\bibfnamefont {J.}~\bibnamefont {Waissman}}, \bibinfo {author}
  {\bibfnamefont {K.}~\bibnamefont {Kaasbjerg}}, \bibinfo {author}
  {\bibfnamefont {Y.}~\bibnamefont {Oreg}}, \bibinfo {author} {\bibfnamefont
  {F.}~\bibnamefont {{Von Oppen}}}, \ and\ \bibinfo {author} {\bibfnamefont
  {S.}~\bibnamefont {Ilani}},\ }\href {\doibase 10.1038/nature18639} {\bibfield
   {journal} {\bibinfo  {journal} {Nature}\ }\textbf {\bibinfo {volume}
  {535}},\ \bibinfo {pages} {395} (\bibinfo {year} {2016})}\BibitemShut
  {NoStop}%
\bibitem [{\citenamefont {Morel}\ and\ \citenamefont
  {Anderson}(1962)}]{Anderson1962}%
  \BibitemOpen
  \bibfield  {author} {\bibinfo {author} {\bibfnamefont {P.}~\bibnamefont
  {Morel}}\ and\ \bibinfo {author} {\bibfnamefont {P.~W.}\ \bibnamefont
  {Anderson}},\ }\href {\doibase 10.1103/PhysRev.125.1263} {\bibfield
  {journal} {\bibinfo  {journal} {Phys. Rev.}\ }\textbf {\bibinfo {volume}
  {125}},\ \bibinfo {pages} {1263} (\bibinfo {year} {1962})}\BibitemShut
  {NoStop}%
\bibitem [{\citenamefont {Geim}\ and\ \citenamefont
  {Grigorieva}(2013)}]{Geim2013van}%
  \BibitemOpen
  \bibfield  {author} {\bibinfo {author} {\bibfnamefont {A.~K.}\ \bibnamefont
  {Geim}}\ and\ \bibinfo {author} {\bibfnamefont {I.~V.}\ \bibnamefont
  {Grigorieva}},\ }\href@noop {} {\bibfield  {journal} {\bibinfo  {journal}
  {Nature}\ }\textbf {\bibinfo {volume} {499}},\ \bibinfo {pages} {419}
  (\bibinfo {year} {2013})}\BibitemShut {NoStop}%
\bibitem [{\citenamefont {Vakili}\ \emph {et~al.}(2004)\citenamefont {Vakili},
  \citenamefont {Shkolnikov}, \citenamefont {Tutuc}, \citenamefont {{De
  Poortere}},\ and\ \citenamefont {Shayegan}}]{Vakili2004}%
  \BibitemOpen
  \bibfield  {author} {\bibinfo {author} {\bibfnamefont {K.}~\bibnamefont
  {Vakili}}, \bibinfo {author} {\bibfnamefont {Y.~P.}\ \bibnamefont
  {Shkolnikov}}, \bibinfo {author} {\bibfnamefont {E.}~\bibnamefont {Tutuc}},
  \bibinfo {author} {\bibfnamefont {E.~P.}\ \bibnamefont {{De Poortere}}}, \
  and\ \bibinfo {author} {\bibfnamefont {M.}~\bibnamefont {Shayegan}},\ }\href
  {\doibase 10.1103/PhysRevLett.92.186404} {\bibfield  {journal} {\bibinfo
  {journal} {Physical Review Letters}\ }\textbf {\bibinfo {volume} {92}},\
  \bibinfo {pages} {90} (\bibinfo {year} {2004})},\ \Eprint
  {http://arxiv.org/abs/0309385v1} {arXiv:0309385v1 [arXiv:cond-mat]}
  \BibitemShut {NoStop}%
\bibitem [{\citenamefont {Thakur}\ \emph {et~al.}(1998)\citenamefont {Thakur},
  \citenamefont {Neilson},\ and\ \citenamefont {Das}}]{thakur1998electron}%
  \BibitemOpen
  \bibfield  {author} {\bibinfo {author} {\bibfnamefont {J.}~\bibnamefont
  {Thakur}}, \bibinfo {author} {\bibfnamefont {D.}~\bibnamefont {Neilson}}, \
  and\ \bibinfo {author} {\bibfnamefont {M.}~\bibnamefont {Das}},\ }\href@noop
  {} {\bibfield  {journal} {\bibinfo  {journal} {Physical Review B}\ }\textbf
  {\bibinfo {volume} {57}},\ \bibinfo {pages} {1801} (\bibinfo {year}
  {1998})}\BibitemShut {NoStop}%
\bibitem [{\citenamefont {Bolotin}\ \emph {et~al.}(2008)\citenamefont
  {Bolotin}, \citenamefont {Sikes}, \citenamefont {Jiang}, \citenamefont
  {Klima}, \citenamefont {Fudenberg}, \citenamefont {Hone}, \citenamefont
  {Kim},\ and\ \citenamefont {Stormer}}]{bolotin2008ultrahigh}%
  \BibitemOpen
  \bibfield  {author} {\bibinfo {author} {\bibfnamefont {K.~I.}\ \bibnamefont
  {Bolotin}}, \bibinfo {author} {\bibfnamefont {K.}~\bibnamefont {Sikes}},
  \bibinfo {author} {\bibfnamefont {Z.}~\bibnamefont {Jiang}}, \bibinfo
  {author} {\bibfnamefont {M.}~\bibnamefont {Klima}}, \bibinfo {author}
  {\bibfnamefont {G.}~\bibnamefont {Fudenberg}}, \bibinfo {author}
  {\bibfnamefont {J.}~\bibnamefont {Hone}}, \bibinfo {author} {\bibfnamefont
  {P.}~\bibnamefont {Kim}}, \ and\ \bibinfo {author} {\bibfnamefont
  {H.}~\bibnamefont {Stormer}},\ }\href@noop {} {\bibfield  {journal} {\bibinfo
   {journal} {Solid State Communications}\ }\textbf {\bibinfo {volume} {146}},\
  \bibinfo {pages} {351} (\bibinfo {year} {2008})}\BibitemShut {NoStop}%
\bibitem [{\citenamefont {Du}\ \emph {et~al.}(2008)\citenamefont {Du},
  \citenamefont {Skachko}, \citenamefont {Barker},\ and\ \citenamefont
  {Andrei}}]{du2008approaching}%
  \BibitemOpen
  \bibfield  {author} {\bibinfo {author} {\bibfnamefont {X.}~\bibnamefont
  {Du}}, \bibinfo {author} {\bibfnamefont {I.}~\bibnamefont {Skachko}},
  \bibinfo {author} {\bibfnamefont {A.}~\bibnamefont {Barker}}, \ and\ \bibinfo
  {author} {\bibfnamefont {E.~Y.}\ \bibnamefont {Andrei}},\ }\href@noop {}
  {\bibfield  {journal} {\bibinfo  {journal} {Nature nanotechnology}\ }\textbf
  {\bibinfo {volume} {3}},\ \bibinfo {pages} {491} (\bibinfo {year}
  {2008})}\BibitemShut {NoStop}%
\bibitem [{\citenamefont {Weitz}\ \emph {et~al.}(2010)\citenamefont {Weitz},
  \citenamefont {Allen}, \citenamefont {Feldman}, \citenamefont {Martin},\ and\
  \citenamefont {Yacoby}}]{weitz2010broken}%
  \BibitemOpen
  \bibfield  {author} {\bibinfo {author} {\bibfnamefont {R.~T.}\ \bibnamefont
  {Weitz}}, \bibinfo {author} {\bibfnamefont {M.~T.}\ \bibnamefont {Allen}},
  \bibinfo {author} {\bibfnamefont {B.~E.}\ \bibnamefont {Feldman}}, \bibinfo
  {author} {\bibfnamefont {J.}~\bibnamefont {Martin}}, \ and\ \bibinfo {author}
  {\bibfnamefont {A.}~\bibnamefont {Yacoby}},\ }\href@noop {} {\bibfield
  {journal} {\bibinfo  {journal} {Science}\ }\textbf {\bibinfo {volume}
  {330}},\ \bibinfo {pages} {812} (\bibinfo {year} {2010})}\BibitemShut
  {NoStop}%
\bibitem [{\citenamefont {Takada}(1980)}]{Takada1980}%
  \BibitemOpen
  \bibfield  {author} {\bibinfo {author} {\bibfnamefont {Y.}~\bibnamefont
  {Takada}},\ }\href {\doibase 10.1143/JPSJ.49.1713} {\bibfield  {journal}
  {\bibinfo  {journal} {Journal of the Physical Society of Japan}\ }\textbf
  {\bibinfo {volume} {49}},\ \bibinfo {pages} {1713} (\bibinfo {year}
  {1980})}\BibitemShut {NoStop}%
\bibitem [{\citenamefont {Pashitskii}(1969)}]{Pashitskii1969}%
  \BibitemOpen
  \bibfield  {author} {\bibinfo {author} {\bibfnamefont {E.~A.}\ \bibnamefont
  {Pashitskii}},\ }\href@noop {} {\bibfield  {journal} {\bibinfo  {journal}
  {Soviet Physics JETP}\ }\textbf {\bibinfo {volume} {28}},\ \bibinfo {pages}
  {1267} (\bibinfo {year} {1969})}\BibitemShut {NoStop}%
\bibitem [{\citenamefont {Fr{\"o}hlich}(1968)}]{frohlich1968superconductivity}%
  \BibitemOpen
  \bibfield  {author} {\bibinfo {author} {\bibfnamefont {H.}~\bibnamefont
  {Fr{\"o}hlich}},\ }\href@noop {} {\bibfield  {journal} {\bibinfo  {journal}
  {Journal of Physics C: Solid State Physics}\ }\textbf {\bibinfo {volume}
  {1}},\ \bibinfo {pages} {544} (\bibinfo {year} {1968})}\BibitemShut {NoStop}%
\bibitem [{\citenamefont {Radhakrishnan}(1965)}]{RADHAKRISHNAN1965247}%
  \BibitemOpen
  \bibfield  {author} {\bibinfo {author} {\bibfnamefont {V.}~\bibnamefont
  {Radhakrishnan}},\ }\href {\doibase
  http://dx.doi.org/10.1016/0031-9163(65)90829-2} {\bibfield  {journal}
  {\bibinfo  {journal} {Physics Letters}\ }\textbf {\bibinfo {volume} {16}},\
  \bibinfo {pages} {247 } (\bibinfo {year} {1965})}\BibitemShut {NoStop}%
\bibitem [{\citenamefont {Entin-Wohlman}\ and\ \citenamefont
  {Gutfreund}(1984)}]{entin1984acoustic}%
  \BibitemOpen
  \bibfield  {author} {\bibinfo {author} {\bibfnamefont {O.}~\bibnamefont
  {Entin-Wohlman}}\ and\ \bibinfo {author} {\bibfnamefont {H.}~\bibnamefont
  {Gutfreund}},\ }\href@noop {} {\bibfield  {journal} {\bibinfo  {journal}
  {Journal of Physics C: Solid State Physics}\ }\textbf {\bibinfo {volume}
  {17}},\ \bibinfo {pages} {1071} (\bibinfo {year} {1984})}\BibitemShut
  {NoStop}%
\bibitem [{\citenamefont {Garland}(1963)}]{Garland}%
  \BibitemOpen
  \bibfield  {author} {\bibinfo {author} {\bibfnamefont {J.~W.}\ \bibnamefont
  {Garland}},\ }\href {\doibase 10.1103/PhysRevLett.11.111} {\bibfield
  {journal} {\bibinfo  {journal} {Phys. Rev. Lett.}\ }\textbf {\bibinfo
  {volume} {11}},\ \bibinfo {pages} {111} (\bibinfo {year} {1963})}\BibitemShut
  {NoStop}%
\bibitem [{\citenamefont {Canright}\ and\ \citenamefont
  {Vignale}(1989)}]{Canright1989}%
  \BibitemOpen
  \bibfield  {author} {\bibinfo {author} {\bibfnamefont {G.~S.}\ \bibnamefont
  {Canright}}\ and\ \bibinfo {author} {\bibfnamefont {G.}~\bibnamefont
  {Vignale}},\ }\href {\doibase 10.1103/PhysRevB.39.2740} {\bibfield  {journal}
  {\bibinfo  {journal} {Physical Review B}\ }\textbf {\bibinfo {volume} {39}},\
  \bibinfo {pages} {2740} (\bibinfo {year} {1989})}\BibitemShut {NoStop}%
\bibitem [{\citenamefont {Uchoa}\ and\ \citenamefont
  {Neto}(2007)}]{uchoa2007superconducting}%
  \BibitemOpen
  \bibfield  {author} {\bibinfo {author} {\bibfnamefont {B.}~\bibnamefont
  {Uchoa}}\ and\ \bibinfo {author} {\bibfnamefont {A.~C.}\ \bibnamefont
  {Neto}},\ }\href@noop {} {\bibfield  {journal} {\bibinfo  {journal} {Physical
  Review Letters}\ }\textbf {\bibinfo {volume} {98}},\ \bibinfo {pages}
  {146801} (\bibinfo {year} {2007})}\BibitemShut {NoStop}%
\bibitem [{\citenamefont {Santoro}\ and\ \citenamefont
  {Giuliani}(1988)}]{Gabriele1988}%
  \BibitemOpen
  \bibfield  {author} {\bibinfo {author} {\bibfnamefont {G.~E.}\ \bibnamefont
  {Santoro}}\ and\ \bibinfo {author} {\bibfnamefont {G.~F.}\ \bibnamefont
  {Giuliani}},\ }\href {\doibase 10.1103/PhysRevB.37.937} {\bibfield  {journal}
  {\bibinfo  {journal} {Phys. Rev. B}\ }\textbf {\bibinfo {volume} {37}},\
  \bibinfo {pages} {937} (\bibinfo {year} {1988})}\BibitemShut {NoStop}%
\bibitem [{\citenamefont {Hwang}\ and\ \citenamefont
  {Das~Sarma}(2009)}]{Hwang2009}%
  \BibitemOpen
  \bibfield  {author} {\bibinfo {author} {\bibfnamefont {E.~H.}\ \bibnamefont
  {Hwang}}\ and\ \bibinfo {author} {\bibfnamefont {S.}~\bibnamefont
  {Das~Sarma}},\ }\href {\doibase 10.1103/PhysRevB.80.205405} {\bibfield
  {journal} {\bibinfo  {journal} {Phys. Rev. B}\ }\textbf {\bibinfo {volume}
  {80}},\ \bibinfo {pages} {205405} (\bibinfo {year} {2009})}\BibitemShut
  {NoStop}%
\bibitem [{\citenamefont {Profumo}\ \emph
  {et~al.}(2012{\natexlab{a}})\citenamefont {Profumo}, \citenamefont {Asgari},
  \citenamefont {Polini},\ and\ \citenamefont {MacDonald}}]{Rosario2012}%
  \BibitemOpen
  \bibfield  {author} {\bibinfo {author} {\bibfnamefont {R.~E.~V.}\
  \bibnamefont {Profumo}}, \bibinfo {author} {\bibfnamefont {R.}~\bibnamefont
  {Asgari}}, \bibinfo {author} {\bibfnamefont {M.}~\bibnamefont {Polini}}, \
  and\ \bibinfo {author} {\bibfnamefont {A.~H.}\ \bibnamefont {MacDonald}},\
  }\href {\doibase 10.1103/PhysRevB.85.085443} {\bibfield  {journal} {\bibinfo
  {journal} {Phys. Rev. B}\ }\textbf {\bibinfo {volume} {85}},\ \bibinfo
  {pages} {085443} (\bibinfo {year} {2012}{\natexlab{a}})}\BibitemShut
  {NoStop}%
\bibitem [{\citenamefont {Gurevich}\ \emph {et~al.}(1962)\citenamefont
  {Gurevich}, \citenamefont {Larkin},\ and\ \citenamefont
  {Firsov}}]{Gurevich1962}%
  \BibitemOpen
  \bibfield  {author} {\bibinfo {author} {\bibfnamefont {L.~V.}\ \bibnamefont
  {Gurevich}}, \bibinfo {author} {\bibfnamefont {A.~I.}\ \bibnamefont
  {Larkin}}, \ and\ \bibinfo {author} {\bibfnamefont {Y.~A.}\ \bibnamefont
  {Firsov}},\ }\href@noop {} {\bibfield  {journal} {\bibinfo  {journal} {Sov.
  Phys. Sol. State}\ }\textbf {\bibinfo {volume} {4}},\ \bibinfo {pages} {131}
  (\bibinfo {year} {1962})}\BibitemShut {NoStop}%
\bibitem [{\citenamefont {Pines}(1956)}]{pines1956electron}%
  \BibitemOpen
  \bibfield  {author} {\bibinfo {author} {\bibfnamefont {D.}~\bibnamefont
  {Pines}},\ }\href@noop {} {\bibfield  {journal} {\bibinfo  {journal}
  {Canadian Journal of Physics}\ }\textbf {\bibinfo {volume} {34}},\ \bibinfo
  {pages} {1379} (\bibinfo {year} {1956})}\BibitemShut {NoStop}%
\bibitem [{\citenamefont {Bennacer}\ and\ \citenamefont
  {Cottey}(1989)}]{bennacer1989acoustic}%
  \BibitemOpen
  \bibfield  {author} {\bibinfo {author} {\bibfnamefont {B.}~\bibnamefont
  {Bennacer}}\ and\ \bibinfo {author} {\bibfnamefont {A.}~\bibnamefont
  {Cottey}},\ }\href
  {http://iopscience.iop.org/article/10.1088/0953-8984/1/10/003/meta}
  {\bibfield  {journal} {\bibinfo  {journal} {Journal of Physics: Condensed
  Matter}\ }\textbf {\bibinfo {volume} {1.10}},\ \bibinfo {pages} {1809}
  (\bibinfo {year} {1989})}\BibitemShut {NoStop}%
\bibitem [{\citenamefont {Bennacer}\ \emph {et~al.}(1989)\citenamefont
  {Bennacer}, \citenamefont {Cottey},\ and\ \citenamefont
  {Senkiw}}]{bennacer1989calculated}%
  \BibitemOpen
  \bibfield  {author} {\bibinfo {author} {\bibfnamefont {B.}~\bibnamefont
  {Bennacer}}, \bibinfo {author} {\bibfnamefont {A.}~\bibnamefont {Cottey}}, \
  and\ \bibinfo {author} {\bibfnamefont {J.}~\bibnamefont {Senkiw}},\ }\href
  {http://iopscience.iop.org/article/10.1088/0953-8984/1/45/013/meta}
  {\bibfield  {journal} {\bibinfo  {journal} {Journal of Physics: Condensed
  Matter}\ }\textbf {\bibinfo {volume} {1.45}},\ \bibinfo {pages} {8877}
  (\bibinfo {year} {1989})}\BibitemShut {NoStop}%
\bibitem [{\citenamefont {Chudzinski}\ and\ \citenamefont
  {Giamarchi}(2011)}]{chudzinski2011collective}%
  \BibitemOpen
  \bibfield  {author} {\bibinfo {author} {\bibfnamefont {P.}~\bibnamefont
  {Chudzinski}}\ and\ \bibinfo {author} {\bibfnamefont {T.}~\bibnamefont
  {Giamarchi}},\ }\href@noop {} {\bibfield  {journal} {\bibinfo  {journal}
  {Physical Review B}\ }\textbf {\bibinfo {volume} {84}},\ \bibinfo {pages}
  {125105} (\bibinfo {year} {2011})}\BibitemShut {NoStop}%
\bibitem [{\citenamefont
  {De~Gennes}(2018{\natexlab{a}})}]{de2018superconductivity}%
  \BibitemOpen
  \bibfield  {author} {\bibinfo {author} {\bibfnamefont {P.-G.}\ \bibnamefont
  {De~Gennes}},\ }\href@noop {} {\emph {\bibinfo {title} {Superconductivity of
  metals and alloys}}}\ (\bibinfo  {publisher} {CRC Press},\ \bibinfo {year}
  {2018})\BibitemShut {NoStop}%
\bibitem [{\citenamefont {Ruhman}\ and\ \citenamefont
  {Lee}(2017)}]{Ruhman2017}%
  \BibitemOpen
  \bibfield  {author} {\bibinfo {author} {\bibfnamefont {J.}~\bibnamefont
  {Ruhman}}\ and\ \bibinfo {author} {\bibfnamefont {P.~A.}\ \bibnamefont
  {Lee}},\ }\href {\doibase 10.1103/PhysRevB.96.235107} {\bibfield  {journal}
  {\bibinfo  {journal} {Phys. Rev. B}\ }\textbf {\bibinfo {volume} {96}},\
  \bibinfo {pages} {235107} (\bibinfo {year} {2017})}\BibitemShut {NoStop}%
\bibitem [{\citenamefont {Margine}\ and\ \citenamefont
  {Giustino}(2013)}]{Margine2013}%
  \BibitemOpen
  \bibfield  {author} {\bibinfo {author} {\bibfnamefont {E.~R.}\ \bibnamefont
  {Margine}}\ and\ \bibinfo {author} {\bibfnamefont {F.}~\bibnamefont
  {Giustino}},\ }\href {\doibase 10.1103/PhysRevB.87.024505} {\bibfield
  {journal} {\bibinfo  {journal} {Phys. Rev. B}\ }\textbf {\bibinfo {volume}
  {87}},\ \bibinfo {pages} {024505} (\bibinfo {year} {2013})}\BibitemShut
  {NoStop}%
\bibitem [{\citenamefont {De~Gennes}(2018{\natexlab{b}})}]{de2018}%
  \BibitemOpen
  \bibfield  {author} {\bibinfo {author} {\bibfnamefont {P.-G.}\ \bibnamefont
  {De~Gennes}},\ }\href@noop {} {\emph {\bibinfo {title} {Superconductivity of
  metals and alloys}}}\ (\bibinfo  {publisher} {CRC Press},\ \bibinfo {year}
  {2018})\BibitemShut {NoStop}%
\bibitem [{\citenamefont {Fallahazad}\ \emph {et~al.}(2016)\citenamefont
  {Fallahazad}, \citenamefont {Movva}, \citenamefont {Kim}, \citenamefont
  {Larentis}, \citenamefont {Taniguchi}, \citenamefont {Watanabe},
  \citenamefont {Banerjee},\ and\ \citenamefont {Tutuc}}]{Babak2016}%
  \BibitemOpen
  \bibfield  {author} {\bibinfo {author} {\bibfnamefont {B.}~\bibnamefont
  {Fallahazad}}, \bibinfo {author} {\bibfnamefont {H.~C.~P.}\ \bibnamefont
  {Movva}}, \bibinfo {author} {\bibfnamefont {K.}~\bibnamefont {Kim}}, \bibinfo
  {author} {\bibfnamefont {S.}~\bibnamefont {Larentis}}, \bibinfo {author}
  {\bibfnamefont {T.}~\bibnamefont {Taniguchi}}, \bibinfo {author}
  {\bibfnamefont {K.}~\bibnamefont {Watanabe}}, \bibinfo {author}
  {\bibfnamefont {S.~K.}\ \bibnamefont {Banerjee}}, \ and\ \bibinfo {author}
  {\bibfnamefont {E.}~\bibnamefont {Tutuc}},\ }\href {\doibase
  10.1103/PhysRevLett.116.086601} {\bibfield  {journal} {\bibinfo  {journal}
  {Phys. Rev. Lett.}\ }\textbf {\bibinfo {volume} {116}},\ \bibinfo {pages}
  {086601} (\bibinfo {year} {2016})}\BibitemShut {NoStop}%
\bibitem [{\citenamefont {Gustafsson}\ \emph {et~al.}(2017)\citenamefont
  {Gustafsson}, \citenamefont {Yankowitz}, \citenamefont {Forsythe},
  \citenamefont {Rhodes}, \citenamefont {Watanabe}, \citenamefont {Taniguchi},
  \citenamefont {Hone}, \citenamefont {Zhu},\ and\ \citenamefont
  {Dean}}]{Gustafsson2017arxiv}%
  \BibitemOpen
  \bibfield  {author} {\bibinfo {author} {\bibfnamefont {M.~V.}\ \bibnamefont
  {Gustafsson}}, \bibinfo {author} {\bibfnamefont {M.}~\bibnamefont
  {Yankowitz}}, \bibinfo {author} {\bibfnamefont {C.}~\bibnamefont {Forsythe}},
  \bibinfo {author} {\bibfnamefont {D.}~\bibnamefont {Rhodes}}, \bibinfo
  {author} {\bibfnamefont {K.}~\bibnamefont {Watanabe}}, \bibinfo {author}
  {\bibfnamefont {T.}~\bibnamefont {Taniguchi}}, \bibinfo {author}
  {\bibfnamefont {J.}~\bibnamefont {Hone}}, \bibinfo {author} {\bibfnamefont
  {X.}~\bibnamefont {Zhu}}, \ and\ \bibinfo {author} {\bibfnamefont {C.~R.}\
  \bibnamefont {Dean}},\ }\href@noop {} {\enquote {\bibinfo {title} {Ambipolar
  landau levels and strong band-selective carrier interactions in monolayer
  wse$_2$},}\ } (\bibinfo {year} {2017}),\ \Eprint
  {http://arxiv.org/abs/arXiv:1707.08083} {arXiv:1707.08083} \BibitemShut
  {NoStop}%
\bibitem [{\citenamefont {Sun}\ \emph {et~al.}(2014)\citenamefont {Sun},
  \citenamefont {Ding}, \citenamefont {Goodman}, \citenamefont {Funke},\ and\
  \citenamefont {Nagpal}}]{Sun2014}%
  \BibitemOpen
  \bibfield  {author} {\bibinfo {author} {\bibfnamefont {Q.-c.}\ \bibnamefont
  {Sun}}, \bibinfo {author} {\bibfnamefont {Y.}~\bibnamefont {Ding}}, \bibinfo
  {author} {\bibfnamefont {S.~M.}\ \bibnamefont {Goodman}}, \bibinfo {author}
  {\bibfnamefont {H.~H.}\ \bibnamefont {Funke}}, \ and\ \bibinfo {author}
  {\bibfnamefont {P.}~\bibnamefont {Nagpal}},\ }\href {\doibase
  10.1039/c4nr04719b} {\bibfield  {journal} {\bibinfo  {journal} {Nanoscale}\
  }\textbf {\bibinfo {volume} {6}},\ \bibinfo {pages} {12450} (\bibinfo {year}
  {2014})}\BibitemShut {NoStop}%
\bibitem [{\citenamefont {Roesner}\ \emph {et~al.}(2018)\citenamefont
  {Roesner}, \citenamefont {Groenewald}, \citenamefont {Schoenhoff},
  \citenamefont {Berges}, \citenamefont {Haas},\ and\ \citenamefont
  {Wehling}}]{roesner2018plasmonic}%
  \BibitemOpen
  \bibfield  {author} {\bibinfo {author} {\bibfnamefont {M.}~\bibnamefont
  {Roesner}}, \bibinfo {author} {\bibfnamefont {R.}~\bibnamefont {Groenewald}},
  \bibinfo {author} {\bibfnamefont {G.}~\bibnamefont {Schoenhoff}}, \bibinfo
  {author} {\bibfnamefont {J.}~\bibnamefont {Berges}}, \bibinfo {author}
  {\bibfnamefont {S.}~\bibnamefont {Haas}}, \ and\ \bibinfo {author}
  {\bibfnamefont {T.}~\bibnamefont {Wehling}},\ }\href@noop {} {\bibfield
  {journal} {\bibinfo  {journal} {arXiv preprint arXiv:1803.04576}\ } (\bibinfo
  {year} {2018})}\BibitemShut {NoStop}%
\bibitem [{\citenamefont {Savary}\ \emph {et~al.}(2017)\citenamefont {Savary},
  \citenamefont {Ruhman}, \citenamefont {Venderbos}, \citenamefont {Fu},\ and\
  \citenamefont {Lee}}]{Savary2017}%
  \BibitemOpen
  \bibfield  {author} {\bibinfo {author} {\bibfnamefont {L.}~\bibnamefont
  {Savary}}, \bibinfo {author} {\bibfnamefont {J.}~\bibnamefont {Ruhman}},
  \bibinfo {author} {\bibfnamefont {J.~W.~F.}\ \bibnamefont {Venderbos}},
  \bibinfo {author} {\bibfnamefont {L.}~\bibnamefont {Fu}}, \ and\ \bibinfo
  {author} {\bibfnamefont {P.~A.}\ \bibnamefont {Lee}},\ }\href {\doibase
  10.1103/PhysRevB.96.214514} {\bibfield  {journal} {\bibinfo  {journal} {Phys.
  Rev. B}\ }\textbf {\bibinfo {volume} {96}},\ \bibinfo {pages} {214514}
  (\bibinfo {year} {2017})}\BibitemShut {NoStop}%
\bibitem [{\citenamefont {Takada}(1992)}]{takada1992plasmon}%
  \BibitemOpen
  \bibfield  {author} {\bibinfo {author} {\bibfnamefont {Y.}~\bibnamefont
  {Takada}},\ }\href {http://journals.jps.jp/doi/abs/10.1143/JPSJ.61.238}
  {\bibfield  {journal} {\bibinfo  {journal} {Journal of the Physical Society
  of Japan}\ }\textbf {\bibinfo {volume} {61}},\ \bibinfo {pages} {238}
  (\bibinfo {year} {1992})}\BibitemShut {NoStop}%
\bibitem [{Note1()}]{Note1}%
  \BibitemOpen
  \bibinfo {note} {The value of $Q_d / k_d = 8.75$ corresponds to suspended
  graphene.}\BibitemShut {Stop}%
\bibitem [{\citenamefont {Ma}\ \emph {et~al.}(2011)\citenamefont {Ma},
  \citenamefont {Dai}, \citenamefont {Guo}, \citenamefont {Niu},\ and\
  \citenamefont {Huang}}]{ma2011graphene}%
  \BibitemOpen
  \bibfield  {author} {\bibinfo {author} {\bibfnamefont {Y.}~\bibnamefont
  {Ma}}, \bibinfo {author} {\bibfnamefont {Y.}~\bibnamefont {Dai}}, \bibinfo
  {author} {\bibfnamefont {M.}~\bibnamefont {Guo}}, \bibinfo {author}
  {\bibfnamefont {C.}~\bibnamefont {Niu}}, \ and\ \bibinfo {author}
  {\bibfnamefont {B.}~\bibnamefont {Huang}},\ }\href@noop {} {\bibfield
  {journal} {\bibinfo  {journal} {Nanoscale}\ }\textbf {\bibinfo {volume}
  {3}},\ \bibinfo {pages} {3883} (\bibinfo {year} {2011})}\BibitemShut
  {NoStop}%
\bibitem [{\citenamefont {Profumo}\ \emph
  {et~al.}(2012{\natexlab{b}})\citenamefont {Profumo}, \citenamefont {Asgari},
  \citenamefont {Polini},\ and\ \citenamefont {MacDonald}}]{Profumo2012}%
  \BibitemOpen
  \bibfield  {author} {\bibinfo {author} {\bibfnamefont {R.~E.~V.}\
  \bibnamefont {Profumo}}, \bibinfo {author} {\bibfnamefont {R.}~\bibnamefont
  {Asgari}}, \bibinfo {author} {\bibfnamefont {M.}~\bibnamefont {Polini}}, \
  and\ \bibinfo {author} {\bibfnamefont {A.~H.}\ \bibnamefont {MacDonald}},\
  }\href {\doibase 10.1103/PhysRevB.85.085443} {\bibfield  {journal} {\bibinfo
  {journal} {Phys. Rev. B}\ }\textbf {\bibinfo {volume} {85}},\ \bibinfo
  {pages} {085443} (\bibinfo {year} {2012}{\natexlab{b}})}\BibitemShut
  {NoStop}%
\bibitem [{\citenamefont {Barlas}\ \emph {et~al.}(2007)\citenamefont {Barlas},
  \citenamefont {Pereg-Barnea}, \citenamefont {Polini}, \citenamefont
  {Asgari},\ and\ \citenamefont {MacDonald}}]{Barlas2007}%
  \BibitemOpen
  \bibfield  {author} {\bibinfo {author} {\bibfnamefont {Y.}~\bibnamefont
  {Barlas}}, \bibinfo {author} {\bibfnamefont {T.}~\bibnamefont
  {Pereg-Barnea}}, \bibinfo {author} {\bibfnamefont {M.}~\bibnamefont
  {Polini}}, \bibinfo {author} {\bibfnamefont {R.}~\bibnamefont {Asgari}}, \
  and\ \bibinfo {author} {\bibfnamefont {A.~H.}\ \bibnamefont {MacDonald}},\
  }\href {\doibase 10.1103/PhysRevLett.98.236601} {\bibfield  {journal}
  {\bibinfo  {journal} {Phys. Rev. Lett.}\ }\textbf {\bibinfo {volume} {98}},\
  \bibinfo {pages} {236601} (\bibinfo {year} {2007})}\BibitemShut {NoStop}%
\bibitem [{\citenamefont {Eliashberg}(1960)}]{Eliashberg1960}%
  \BibitemOpen
  \bibfield  {author} {\bibinfo {author} {\bibfnamefont {G.~M.}\ \bibnamefont
  {Eliashberg}},\ }\href@noop {} {\bibfield  {journal} {\bibinfo  {journal}
  {Sov. Phys. Sol. JETP}\ }\textbf {\bibinfo {volume} {11}},\ \bibinfo {pages}
  {696} (\bibinfo {year} {1960})}\BibitemShut {NoStop}%
\bibitem [{\citenamefont {Takada}(1978)}]{Takada1978}%
  \BibitemOpen
  \bibfield  {author} {\bibinfo {author} {\bibfnamefont {Y.}~\bibnamefont
  {Takada}},\ }\href {\doibase 10.1143/JPSJ.45.786} {\bibfield  {journal}
  {\bibinfo  {journal} {Journal of the Physical Society of Japan}\ }\textbf
  {\bibinfo {volume} {45}},\ \bibinfo {pages} {786} (\bibinfo {year}
  {1978})}\BibitemShut {NoStop}%
\bibitem [{\citenamefont {Diaconescu}\ \emph {et~al.}(2007)\citenamefont
  {Diaconescu}, \citenamefont {Pohl}, \citenamefont {Vattuone}, \citenamefont
  {Savio}, \citenamefont {Hofmann}, \citenamefont {Silkin}, \citenamefont
  {Pitarke}, \citenamefont {Chulkov}, \citenamefont {Echenique}, \citenamefont
  {Farias} \emph {et~al.}}]{diaconescu2007low}%
  \BibitemOpen
  \bibfield  {author} {\bibinfo {author} {\bibfnamefont {B.}~\bibnamefont
  {Diaconescu}}, \bibinfo {author} {\bibfnamefont {K.}~\bibnamefont {Pohl}},
  \bibinfo {author} {\bibfnamefont {L.}~\bibnamefont {Vattuone}}, \bibinfo
  {author} {\bibfnamefont {L.}~\bibnamefont {Savio}}, \bibinfo {author}
  {\bibfnamefont {P.}~\bibnamefont {Hofmann}}, \bibinfo {author} {\bibfnamefont
  {V.~M.}\ \bibnamefont {Silkin}}, \bibinfo {author} {\bibfnamefont {J.~M.}\
  \bibnamefont {Pitarke}}, \bibinfo {author} {\bibfnamefont {E.~V.}\
  \bibnamefont {Chulkov}}, \bibinfo {author} {\bibfnamefont {P.~M.}\
  \bibnamefont {Echenique}}, \bibinfo {author} {\bibfnamefont {D.}~\bibnamefont
  {Farias}},  \emph {et~al.},\ }\href@noop {} {\bibfield  {journal} {\bibinfo
  {journal} {Nature}\ }\textbf {\bibinfo {volume} {448}},\ \bibinfo {pages}
  {57} (\bibinfo {year} {2007})}\BibitemShut {NoStop}%
\bibitem [{\citenamefont {Lundeberg}\ \emph {et~al.}(2017)\citenamefont
  {Lundeberg}, \citenamefont {Gao}, \citenamefont {Asgari}, \citenamefont
  {Tan}, \citenamefont {Van~Duppen}, \citenamefont {Autore}, \citenamefont
  {Alonso-Gonz{\'a}lez}, \citenamefont {Woessner}, \citenamefont {Watanabe},
  \citenamefont {Taniguchi} \emph {et~al.}}]{lundeberg2017tuning}%
  \BibitemOpen
  \bibfield  {author} {\bibinfo {author} {\bibfnamefont {M.~B.}\ \bibnamefont
  {Lundeberg}}, \bibinfo {author} {\bibfnamefont {Y.}~\bibnamefont {Gao}},
  \bibinfo {author} {\bibfnamefont {R.}~\bibnamefont {Asgari}}, \bibinfo
  {author} {\bibfnamefont {C.}~\bibnamefont {Tan}}, \bibinfo {author}
  {\bibfnamefont {B.}~\bibnamefont {Van~Duppen}}, \bibinfo {author}
  {\bibfnamefont {M.}~\bibnamefont {Autore}}, \bibinfo {author} {\bibfnamefont
  {P.}~\bibnamefont {Alonso-Gonz{\'a}lez}}, \bibinfo {author} {\bibfnamefont
  {A.}~\bibnamefont {Woessner}}, \bibinfo {author} {\bibfnamefont
  {K.}~\bibnamefont {Watanabe}}, \bibinfo {author} {\bibfnamefont
  {T.}~\bibnamefont {Taniguchi}},  \emph {et~al.},\ }\href@noop {} {\bibfield
  {journal} {\bibinfo  {journal} {Science}\ ,\ \bibinfo {pages} {eaan2735}}
  (\bibinfo {year} {2017})}\BibitemShut {NoStop}%
\bibitem [{\citenamefont {Padmanabhan}\ \emph {et~al.}(2014)\citenamefont
  {Padmanabhan}, \citenamefont {Young}, \citenamefont {Henstridge},
  \citenamefont {Bhowmick}, \citenamefont {Bhattacharya},\ and\ \citenamefont
  {Merlin}}]{Padmanabhan2014}%
  \BibitemOpen
  \bibfield  {author} {\bibinfo {author} {\bibfnamefont {P.}~\bibnamefont
  {Padmanabhan}}, \bibinfo {author} {\bibfnamefont {S.~M.}\ \bibnamefont
  {Young}}, \bibinfo {author} {\bibfnamefont {M.}~\bibnamefont {Henstridge}},
  \bibinfo {author} {\bibfnamefont {S.}~\bibnamefont {Bhowmick}}, \bibinfo
  {author} {\bibfnamefont {P.~K.}\ \bibnamefont {Bhattacharya}}, \ and\
  \bibinfo {author} {\bibfnamefont {R.}~\bibnamefont {Merlin}},\ }\href
  {\doibase 10.1103/PhysRevLett.113.027402} {\bibfield  {journal} {\bibinfo
  {journal} {Phys. Rev. Lett.}\ }\textbf {\bibinfo {volume} {113}},\ \bibinfo
  {pages} {027402} (\bibinfo {year} {2014})}\BibitemShut {NoStop}%
\bibitem [{\citenamefont {Flensberg}\ and\ \citenamefont
  {Hu}(1994)}]{Flensberg1994}%
  \BibitemOpen
  \bibfield  {author} {\bibinfo {author} {\bibfnamefont {K.}~\bibnamefont
  {Flensberg}}\ and\ \bibinfo {author} {\bibfnamefont {B.~Y.-K.}\ \bibnamefont
  {Hu}},\ }\href {\doibase 10.1103/PhysRevLett.73.3572} {\bibfield  {journal}
  {\bibinfo  {journal} {Phys. Rev. Lett.}\ }\textbf {\bibinfo {volume} {73}},\
  \bibinfo {pages} {3572} (\bibinfo {year} {1994})}\BibitemShut {NoStop}%
\bibitem [{\citenamefont {de~Vega}\ and\ \citenamefont {García~de
  Abajo}(2017)}]{de2017plasmon}%
  \BibitemOpen
  \bibfield  {author} {\bibinfo {author} {\bibfnamefont {S.}~\bibnamefont
  {de~Vega}}\ and\ \bibinfo {author} {\bibfnamefont {F.~J.}\ \bibnamefont
  {García~de Abajo}},\ }\href@noop {} {\bibfield  {journal} {\bibinfo
  {journal} {ACS Photonics}\ }\textbf {\bibinfo {volume} {4}},\ \bibinfo
  {pages} {2367} (\bibinfo {year} {2017})}\BibitemShut {NoStop}%
\bibitem [{\citenamefont {Jang}\ \emph {et~al.}(2017)\citenamefont {Jang},
  \citenamefont {Yoo}, \citenamefont {Pfeiffer}, \citenamefont {West},
  \citenamefont {Baldwin},\ and\ \citenamefont {Ashoori}}]{jang2017full}%
  \BibitemOpen
  \bibfield  {author} {\bibinfo {author} {\bibfnamefont {J.}~\bibnamefont
  {Jang}}, \bibinfo {author} {\bibfnamefont {H.~M.}\ \bibnamefont {Yoo}},
  \bibinfo {author} {\bibfnamefont {L.}~\bibnamefont {Pfeiffer}}, \bibinfo
  {author} {\bibfnamefont {K.}~\bibnamefont {West}}, \bibinfo {author}
  {\bibfnamefont {K.}~\bibnamefont {Baldwin}}, \ and\ \bibinfo {author}
  {\bibfnamefont {R.~C.}\ \bibnamefont {Ashoori}},\ }\href@noop {} {\bibfield
  {journal} {\bibinfo  {journal} {Science}\ }\textbf {\bibinfo {volume}
  {358}},\ \bibinfo {pages} {901} (\bibinfo {year} {2017})}\BibitemShut
  {NoStop}%
\bibitem [{\citenamefont {Dvir}\ \emph {et~al.}(2018)\citenamefont {Dvir},
  \citenamefont {Massee}, \citenamefont {Attias}, \citenamefont {Khodas},
  \citenamefont {Aprili}, \citenamefont {Quay},\ and\ \citenamefont
  {Steinberg}}]{dvir2018spectroscopy}%
  \BibitemOpen
  \bibfield  {author} {\bibinfo {author} {\bibfnamefont {T.}~\bibnamefont
  {Dvir}}, \bibinfo {author} {\bibfnamefont {F.}~\bibnamefont {Massee}},
  \bibinfo {author} {\bibfnamefont {L.}~\bibnamefont {Attias}}, \bibinfo
  {author} {\bibfnamefont {M.}~\bibnamefont {Khodas}}, \bibinfo {author}
  {\bibfnamefont {M.}~\bibnamefont {Aprili}}, \bibinfo {author} {\bibfnamefont
  {C.~H.}\ \bibnamefont {Quay}}, \ and\ \bibinfo {author} {\bibfnamefont
  {H.}~\bibnamefont {Steinberg}},\ }\href@noop {} {\bibfield  {journal}
  {\bibinfo  {journal} {Nature Communications}\ }\textbf {\bibinfo {volume}
  {9}},\ \bibinfo {pages} {598} (\bibinfo {year} {2018})}\BibitemShut {NoStop}%
\bibitem [{\citenamefont {Fei}\ \emph {et~al.}(2012)\citenamefont {Fei},
  \citenamefont {Rodin}, \citenamefont {Andreev}, \citenamefont {Bao},
  \citenamefont {McLeod}, \citenamefont {Wagner}, \citenamefont {Zhang},
  \citenamefont {Zhao}, \citenamefont {Thiemens}, \citenamefont {Dominguez}
  \emph {et~al.}}]{fei2012gate}%
  \BibitemOpen
  \bibfield  {author} {\bibinfo {author} {\bibfnamefont {Z.}~\bibnamefont
  {Fei}}, \bibinfo {author} {\bibfnamefont {A.}~\bibnamefont {Rodin}}, \bibinfo
  {author} {\bibfnamefont {G.}~\bibnamefont {Andreev}}, \bibinfo {author}
  {\bibfnamefont {W.}~\bibnamefont {Bao}}, \bibinfo {author} {\bibfnamefont
  {A.}~\bibnamefont {McLeod}}, \bibinfo {author} {\bibfnamefont
  {M.}~\bibnamefont {Wagner}}, \bibinfo {author} {\bibfnamefont
  {L.}~\bibnamefont {Zhang}}, \bibinfo {author} {\bibfnamefont
  {Z.}~\bibnamefont {Zhao}}, \bibinfo {author} {\bibfnamefont {M.}~\bibnamefont
  {Thiemens}}, \bibinfo {author} {\bibfnamefont {G.}~\bibnamefont {Dominguez}},
   \emph {et~al.},\ }\href@noop {} {\bibfield  {journal} {\bibinfo  {journal}
  {Nature}\ }\textbf {\bibinfo {volume} {487}},\ \bibinfo {pages} {82}
  (\bibinfo {year} {2012})}\BibitemShut {NoStop}%
\bibitem [{\citenamefont {Alonso-Gonz{\'a}lez}\ \emph
  {et~al.}(2016)\citenamefont {Alonso-Gonz{\'a}lez}, \citenamefont {Nikitin},
  \citenamefont {Gao}, \citenamefont {Woessner}, \citenamefont {Lundeberg},
  \citenamefont {Principi}, \citenamefont {Forcellini}, \citenamefont {Yan},
  \citenamefont {V{\'e}lez}, \citenamefont {Huber} \emph
  {et~al.}}]{alonso2016acoustic}%
  \BibitemOpen
  \bibfield  {author} {\bibinfo {author} {\bibfnamefont {P.}~\bibnamefont
  {Alonso-Gonz{\'a}lez}}, \bibinfo {author} {\bibfnamefont {A.~Y.}\
  \bibnamefont {Nikitin}}, \bibinfo {author} {\bibfnamefont {Y.}~\bibnamefont
  {Gao}}, \bibinfo {author} {\bibfnamefont {A.}~\bibnamefont {Woessner}},
  \bibinfo {author} {\bibfnamefont {M.~B.}\ \bibnamefont {Lundeberg}}, \bibinfo
  {author} {\bibfnamefont {A.}~\bibnamefont {Principi}}, \bibinfo {author}
  {\bibfnamefont {N.}~\bibnamefont {Forcellini}}, \bibinfo {author}
  {\bibfnamefont {W.}~\bibnamefont {Yan}}, \bibinfo {author} {\bibfnamefont
  {S.}~\bibnamefont {V{\'e}lez}}, \bibinfo {author} {\bibfnamefont {A.~J.}\
  \bibnamefont {Huber}},  \emph {et~al.},\ }\href@noop {} {\bibfield  {journal}
  {\bibinfo  {journal} {Nature nanotechnology}\ } (\bibinfo {year}
  {2016})}\BibitemShut {NoStop}%
\bibitem [{\citenamefont {Bhatti}\ \emph {et~al.}(1996)\citenamefont {Bhatti},
  \citenamefont {Richards}, \citenamefont {Hughes},\ and\ \citenamefont
  {Ritchie}}]{PhysRevB.53.11016}%
  \BibitemOpen
  \bibfield  {author} {\bibinfo {author} {\bibfnamefont {A.~S.}\ \bibnamefont
  {Bhatti}}, \bibinfo {author} {\bibfnamefont {D.}~\bibnamefont {Richards}},
  \bibinfo {author} {\bibfnamefont {H.~P.}\ \bibnamefont {Hughes}}, \ and\
  \bibinfo {author} {\bibfnamefont {D.~A.}\ \bibnamefont {Ritchie}},\ }\href
  {\doibase 10.1103/PhysRevB.53.11016} {\bibfield  {journal} {\bibinfo
  {journal} {Phys. Rev. B}\ }\textbf {\bibinfo {volume} {53}},\ \bibinfo
  {pages} {11016} (\bibinfo {year} {1996})}\BibitemShut {NoStop}%
\bibitem [{\citenamefont {Kainth}\ \emph {et~al.}(1998)\citenamefont {Kainth},
  \citenamefont {Richards}, \citenamefont {Hughes}, \citenamefont {Simmons},\
  and\ \citenamefont {Ritchie}}]{PhysRevB.57.R2065}%
  \BibitemOpen
  \bibfield  {author} {\bibinfo {author} {\bibfnamefont {D.~S.}\ \bibnamefont
  {Kainth}}, \bibinfo {author} {\bibfnamefont {D.}~\bibnamefont {Richards}},
  \bibinfo {author} {\bibfnamefont {H.~P.}\ \bibnamefont {Hughes}}, \bibinfo
  {author} {\bibfnamefont {M.~Y.}\ \bibnamefont {Simmons}}, \ and\ \bibinfo
  {author} {\bibfnamefont {D.~A.}\ \bibnamefont {Ritchie}},\ }\href {\doibase
  10.1103/PhysRevB.57.R2065} {\bibfield  {journal} {\bibinfo  {journal} {Phys.
  Rev. B}\ }\textbf {\bibinfo {volume} {57}},\ \bibinfo {pages} {R2065}
  (\bibinfo {year} {1998})}\BibitemShut {NoStop}%
\bibitem [{\citenamefont {Kainth}\ \emph {et~al.}(1999)\citenamefont {Kainth},
  \citenamefont {Richards}, \citenamefont {Bhatti}, \citenamefont {Hughes},
  \citenamefont {Simmons}, \citenamefont {Linfield},\ and\ \citenamefont
  {Ritchie}}]{PhysRevB.59.2095}%
  \BibitemOpen
  \bibfield  {author} {\bibinfo {author} {\bibfnamefont {D.~S.}\ \bibnamefont
  {Kainth}}, \bibinfo {author} {\bibfnamefont {D.}~\bibnamefont {Richards}},
  \bibinfo {author} {\bibfnamefont {A.~S.}\ \bibnamefont {Bhatti}}, \bibinfo
  {author} {\bibfnamefont {H.~P.}\ \bibnamefont {Hughes}}, \bibinfo {author}
  {\bibfnamefont {M.~Y.}\ \bibnamefont {Simmons}}, \bibinfo {author}
  {\bibfnamefont {E.~H.}\ \bibnamefont {Linfield}}, \ and\ \bibinfo {author}
  {\bibfnamefont {D.~A.}\ \bibnamefont {Ritchie}},\ }\href {\doibase
  10.1103/PhysRevB.59.2095} {\bibfield  {journal} {\bibinfo  {journal} {Phys.
  Rev. B}\ }\textbf {\bibinfo {volume} {59}},\ \bibinfo {pages} {2095}
  (\bibinfo {year} {1999})}\BibitemShut {NoStop}%
\end{thebibliography}
\end{document}